\documentclass{aastex631}

\usepackage{graphicx}
\usepackage{txfonts}
\shorttitle{Turbulent cascade in a coronal mass ejection}
\shortauthors{Sorriso-Valvo et al.}
\graphicspath{{./}{figures/}}

\begin{document}

\title{Turbulent cascade and energy transfer rate in a solar coronal mass ejection}

\author{Luca Sorriso-Valvo}
\affiliation{Swedish Institute of Space Physics (IRF), \r{A}ngstr\"om Laboratory, Lägerhyddsvägen 1, SE-751 21 Uppsala, Sweden}
\affiliation{CNR/ISTP -- Istituto per la Scienza e Tecnologia dei Plasmi, Via Amendola 122/D, 70126 Bari, Italy}

\author{Emiliya Yordanova}
\affiliation{Swedish Institute of Space Physics (IRF), \r{A}ngstr\"om Laboratory, Lägerhyddsvägen 1, SE-751 21 Uppsala, Sweden}

\author{Andrew P. Dimmock}
\affiliation{Swedish Institute of Space Physics (IRF), \r{A}ngstr\"om Laboratory, Lägerhyddsvägen 1, SE-751 21 Uppsala, Sweden}

\author{Daniele Telloni}
\affiliation{National Institute for Astrophysics - Astrophysical Observatory of Torino, Via Osservatorio 20, I-10025 Pino Torinese, Italy}



\begin{abstract}


Turbulence properties are examined before, during and after a coronal mass ejection (CME) detected by the Wind spacecraft on July 2012. The power-law scaling of the structure functions, providing information on the power spectral density and flatness of the velocity, magnetic filed and density fluctuations, were examined. The third-order moment scaling law for incompressible, isotropic magnetohydrodynamic turbulence was observed in the preceding and trailing solar wind, as well as in the CME sheath and magnetic cloud. This suggests that the turbulence could develop sufficiently after the shock, or that turbulence in the sheath and cloud regions was robustly preserved even during the mixing with the solar wind plasma.
The turbulent energy transfer rate was thus evaluated in each of the regions. The CME sheath shows an increase of energy transfer rate, as expected from the lower level of Alfv\'enic fluctuations and suggesting the role of the shock-wind interaction as an additional source of energy for the turbulent cascade.

\end{abstract}




\section{Introduction} 
\label{sec:intro}

The interplanetary space is permeated by the solar wind, a rarefied, magnetized plasma that expands from the solar corona  with supersonic and super-Alfv\'enic speed, forming the heliosphere.  
During its expansion, the solar wind interacts with planets and satellites, as well as with their magnetospheres. 
Fast coronal mass ejections~\citep[CMEs,][]{Kilpua2017}, produced by impulsive solar and coronal phenomena, may occasionally impact and perturb the terrestrial magnetosphere, generating variations in the global magnetospheric current system and configuration, and eventually resulting in geomagnetic storms~\citep{Pulkkinen2007}.
Due to their complex structure, impulsive nature and  nonlinear interaction with the embedding turbulent solar wind, the dynamics and evolution of CMEs is difficult to model. Consequently, the CME time of arrival and characteristics at their impact on the magnetosphere are hard to predict, because heliospheric propagation models assume stationary interplanetary medium~\citep{Riley2018}. 
Studies show that solar wind turbulence~\citep{BrunoCarbone2013} is generally strongly developed~\citep{Weygand2007}, anisotropic~\citep{Horbury2012,Narita2018}, intermittent~\citep{Sorriso-Valvo1999}, variable according to the solar wind conditions~\citep{MarschTu1997}, and responsible for the cross-scale transfer of energy that drives solar wind kinetic processes~\citep{Sorriso-Valvo2019,Verscharen2019,Mattheaus2020}, eventually resulting in plasma heating and particle energization~\citep{Vasquez2007,Marino2008,Osman2012,Chasapis2015}. 
More recently, the turbulence properties of CMEs were studied, suggesting that CME-driven sheaths are generally characterized by ``stronger'' turbulence than the preceding solar wind, including larger fluctuation power, steeper spectra, enhanced compressibility and stronger intermittency~\citep{Kilpua2020,Pitna2021}. 
This was ascribed to the dynamical evolution of the CME sheath, where the plasma is compressed and processed at the CME shock, so that power is injected in the system resulting in stronger but ``younger'' or less developed turbulence~\citep{Kilpua2020}. 

In this Letter, we report on the turbulent properties of a fast CME event and of the preceding and trailing solar wind, including the validation of the mixed third-order moment scaling law, and provide the first estimate of the turbulent energy transfer rate in a CME~\citep{PolitanoPouquet1998}. 


\section{Description of data} 
\label{sec:data}

Between 14 and 16 July 2012 an Earth-directed fast CME has been registered by the instrumentation on-board the Wind spacecraft at L1. This event has been previously studied in depth, because its propagation could be followed in detail from the solar origins to L1~\citep{Webb2017}. The CME magnetic cloud was characterized by prolonged duration of the southward magnetic field component that triggered intense geomagnetic storm ~\citep{Mostl2014,Lugaz2016,Hu2016,Scolini2019}. 
The textbook-clear features of the different parts of the CME~\citep[shock, sheath and magnetic cloud,][]{Webb2017} make it suitable for comparing the turbulence properties of the different plasma regions.

Plasma moments and magnetic field measured by the Wind spacecraft will be used to describe the turbulence properties of the solar wind preceding and behind the CME, and of the regions inside the CME. 
The proton velocity $\mathbf{v}$, number density $n_p$ and temperature $T_p$ were measured by the 3DP instrument with 3 s cadence~\citep{Lin1995}. 
The magnetic field $\mathbf{B}$ was measured by the Magnetic Field Investigation magnetometer with 11 Hz cadence~\citep{Lepping1995}, and resampled to the plasma cadence. 
The magnetic field was converted to Alfv\'en velocity units $\mathbf{b}=\mathbf{B}/\sqrt{4\pi n_p m_p}$, where $m_p$ is the proton mass, and the Elsasser variables $\mathbf{z^\pm}=\mathbf{v} \pm \mathbf{b}$ were computed, in order to evaluate the Alfv\'enic properties of the turbulence. 

\begin{figure}
    \centering
    \includegraphics[width=0.8\textwidth]{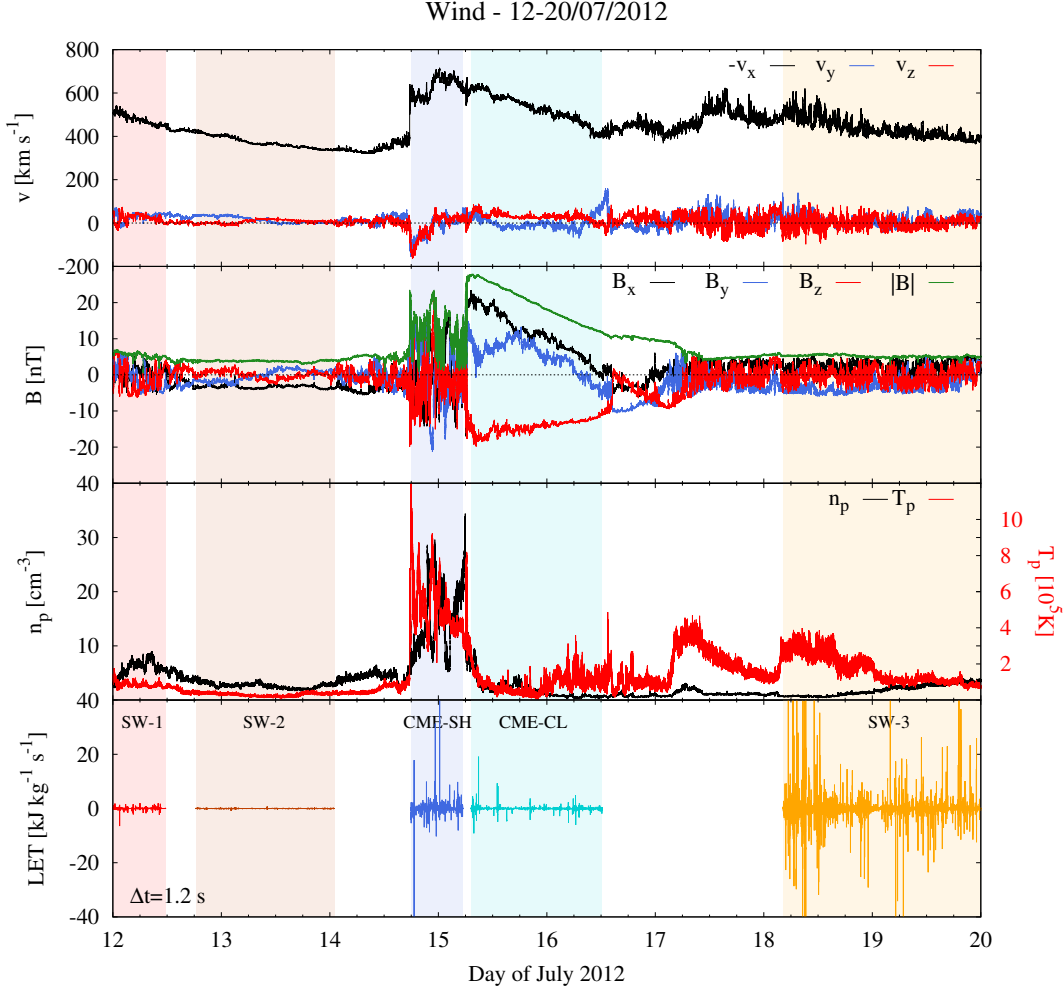}
    \caption{From top to bottom: solar wind velocity components, magnetic field components and magnitude, proton density and temperature, and LET (color-coded for different intervals), for the whole interval. 
    Shaded areas identify the selected sub-intervals, color-coded and labeled in the bottom panel. 
    Vectors are in the GSE coordinate system.}
    \label{fig:data}
\end{figure}

The interval is shown in Figure~\ref{fig:data}. Vectors are in the GSE coordinate system (the $x$ axis pointing from the Earth to the Sun, and the $z$ axis to the ecliptic North). 
The top panel shows the three velocity components, with the dominating $x$ component reversed for clarity.
The second panel shows the three components and magnitude of the magnetic field. 
The third panel shows the proton density and temperature. The fourth panel will be described in Section~\ref{sec:yaglom}.
A few bad data due to plasma instrumental errors have been manually removed before performing the statistical analysis.

From the time series, five sub-intervals were identified as reasonably homogeneous portions within different types of plasma: (i) a pristine solar wind sample preceding the CME, labeled as SW-1 and color-coded in red , with mean solar wind speed $V_{sw}=422$ km/s, and mean kinetic to magnetic pressure ratio $\beta_p=p/p_{mag}=0.53$; (ii) a solar wind sample preceding the CME (SW-2, brown, $V_{sw}=480$ km/s, $\beta_p=10.8$), possibly barring the characteristics of a fast stream rarefaction region, or an interplanetary magnetic flux tube; (iii) a sample in the CME sheath region (CME-SH, blue, $V_{sw}=629$ km/s, $\beta_p=36.6$), where the interplanetary shock arrives at 17:38 UTC on 14 July 2012, followed by intense turbulent fluctuations in the plasma parameters and the magnetic field; (iv) a sample in the CME cloud (CME-CL, turquoise, $V_{sw}=529$ km/s, $\beta_p=0.15$), characterized by smooth magnetic field magnitude, rotation in two of the magnetic components, and low density and temperature; and (v) a sample in the trailing solar wind (SW-3, orange, $V_{sw}=436$ km/s, $\beta_p=0.72$).
Observing an extended period centered on the event under study (not shown), it appears clear that the three solar wind samples belong to different regions, and therefore display the natural variability of solar wind conditions. In particular, SW1 seems to include the slower trailing portion of a preceding fast stream, culminating in the relatively slow, quiet and rarefied SW2 sample. On the other hand, the high level of fluctuations observed in SW3 could be due to fast solar wind behind the CME reaching and compressing the slower trailing part of the cloud region.
All samples are sufficiently large to provide statistical accuracy. 
%
%


\section{Turbulence and intermittency: structure-function analysis} 
\label{sec:turbulence}

In order to evaluate the turbulence properties of each sub-interval, we have used standard analysis tools based on the structure functions, the scale-dependent $q$-th order moments $S_q(\Delta t) = \langle |\Delta \phi^q| \rangle$ of the two-point increments $\Delta \phi = \phi (t+\Delta t) - \phi (t)$ of a field component or scalar $\phi$. 
The structure functions carry information about the scale-dependent statistical properties of the field fluctuations, and are directly related to the autocorrelation function and, therefore, to the energy spectrum~\citep{Frisch1995,DudokDeWit2013}. 
All fields were detrended using a four-hour running average to remove large-scale trends, visible in most of the intervals. 
We have tested that different window sizes do not affect the results. 
In each interval, some extreme values of the fluctuations were removed according to the standard procedure for structure function statistical  convergence~\citep{Kiyani2006}. 

\begin{figure}
    \centering
    \includegraphics[width=0.5\textwidth]{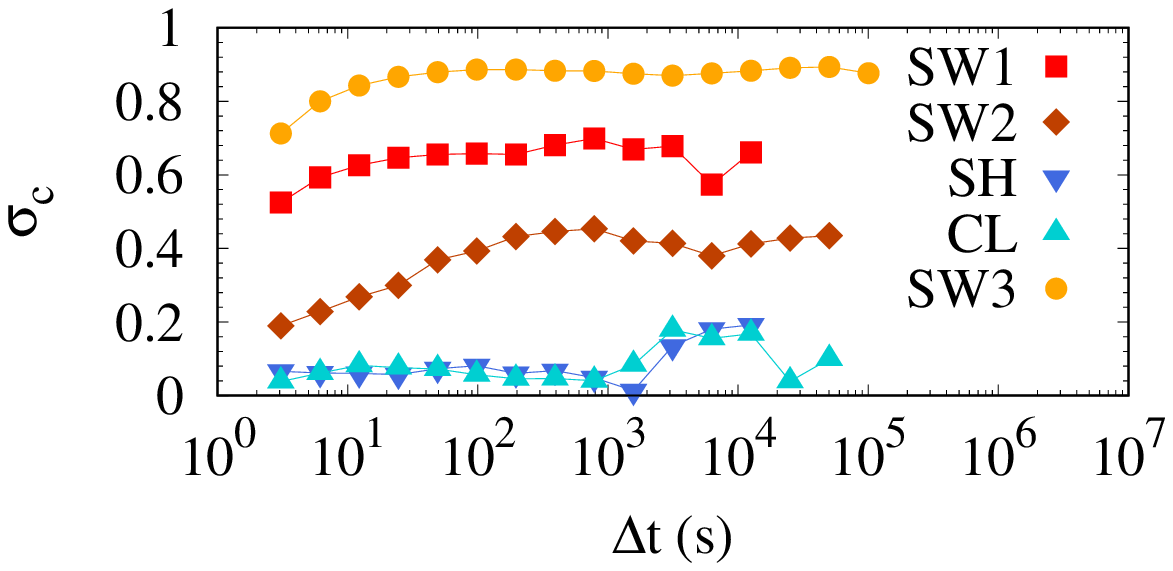}\includegraphics[width=0.5\textwidth]{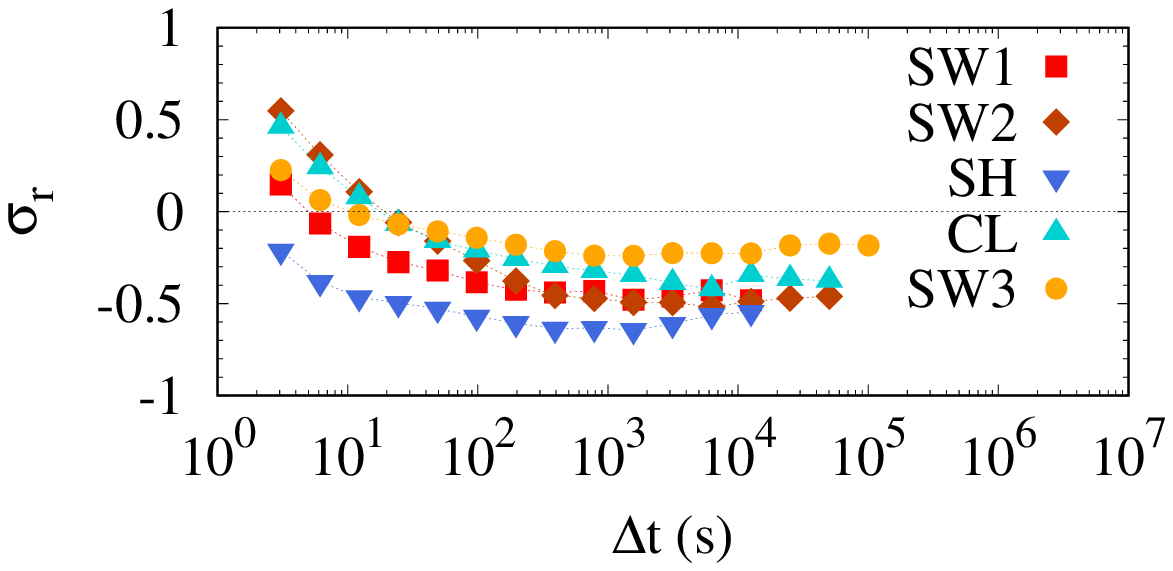}
    \caption{Structure-function-based normalized cross-helicity $\sigma_c$ (left panel) and normalized residual energy $\sigma_r$ (right panel) as a function of scale, for all five regions. All symbols are color-coded for different intervals, according to the labels in the bottom panel of Figure~\ref{fig:data}.}
    \label{fig:sigma}
\end{figure}

The first basic analysis concerns the Alfv\'enic properties of the fluctuations. These can be evaluated using the structure-function-based normalized cross helicity $\sigma_c =[S_2(z^+)-S_2(z^-))]/[S_2(z^+)+S_2(z^-)]$ and residual energy $\sigma_r =[S_2(v)-S_2(b)]/[S_2(v)+S_2(b)]$, where the arguments indicate the trace of the respective vectors~\citep{BrunoCarbone2013}.
These two parameters are plotted in Figure~\ref{fig:sigma} as a function of the scale $\Delta t$ for the five intervals. 
The two samples inside the CME (SH and CL) have nearly zero normalized cross-helicity. This is expected, since, because of its closed-loop field structure, the magnetic cloud is populated by both sunward and antisunward Alfv\'enic fluctuations~\citep{Telloni2020}. The trailing SW3 sample is strongly Alfv\'enic ($\sigma_c \simeq 0.9$), while the preceding SW1 and SW2 samples have moderate Alfv\'enicity ($\sigma_c \simeq 0.6$ and $0.4$, respectively). 
For scales below one minute, the cross-helicity decreases as usual, showing that the turbulent cascade balances the outward and inward Alfv\'enic fluctuations~\citep{DMV1980,BrunoCarbone2013}. 
The normalized residual energy is negative for all samples at most of the scales, and spans from moderate magnetic dominance in the CME sheath region, to nearly balanced turbulence in the trailing solar wind region and in the CME magnetic cloud.
%

\begin{figure}
    \centering
    \includegraphics[width=0.33\textwidth]{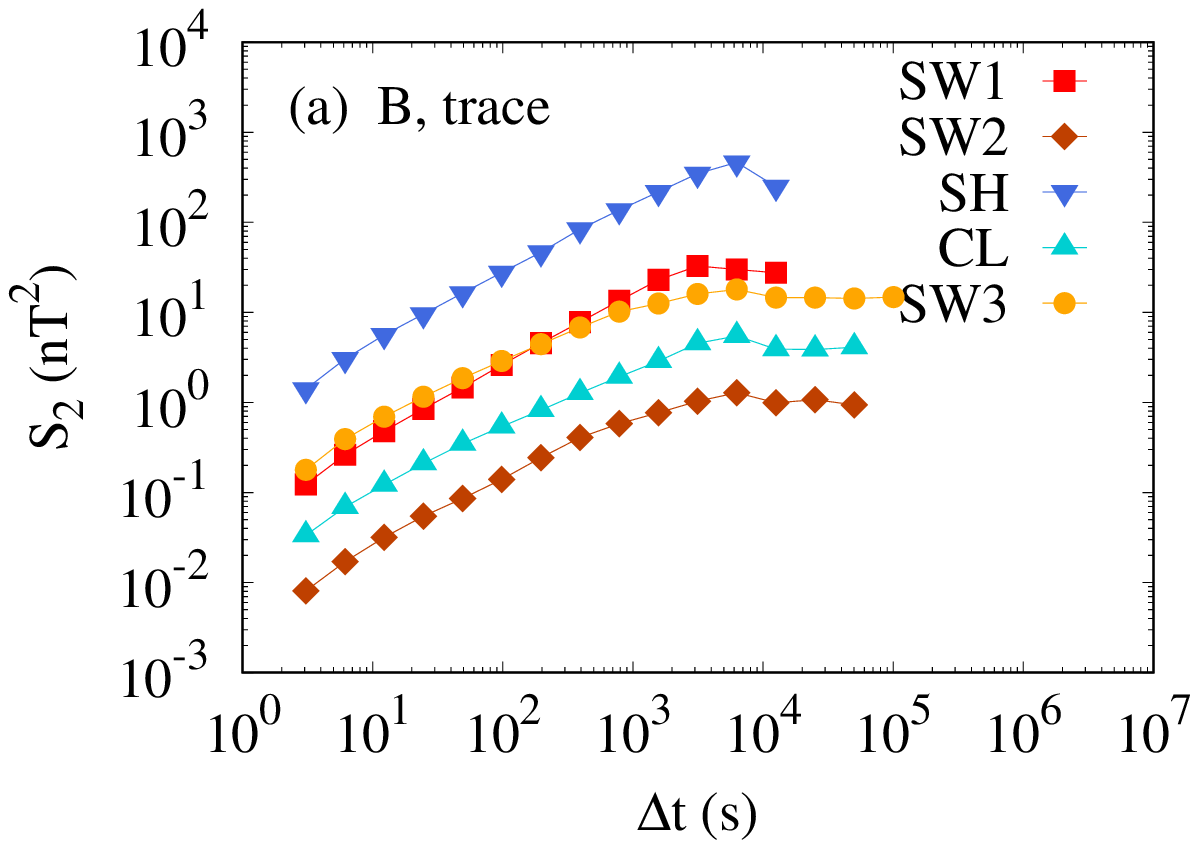}\includegraphics[width=0.33\textwidth]{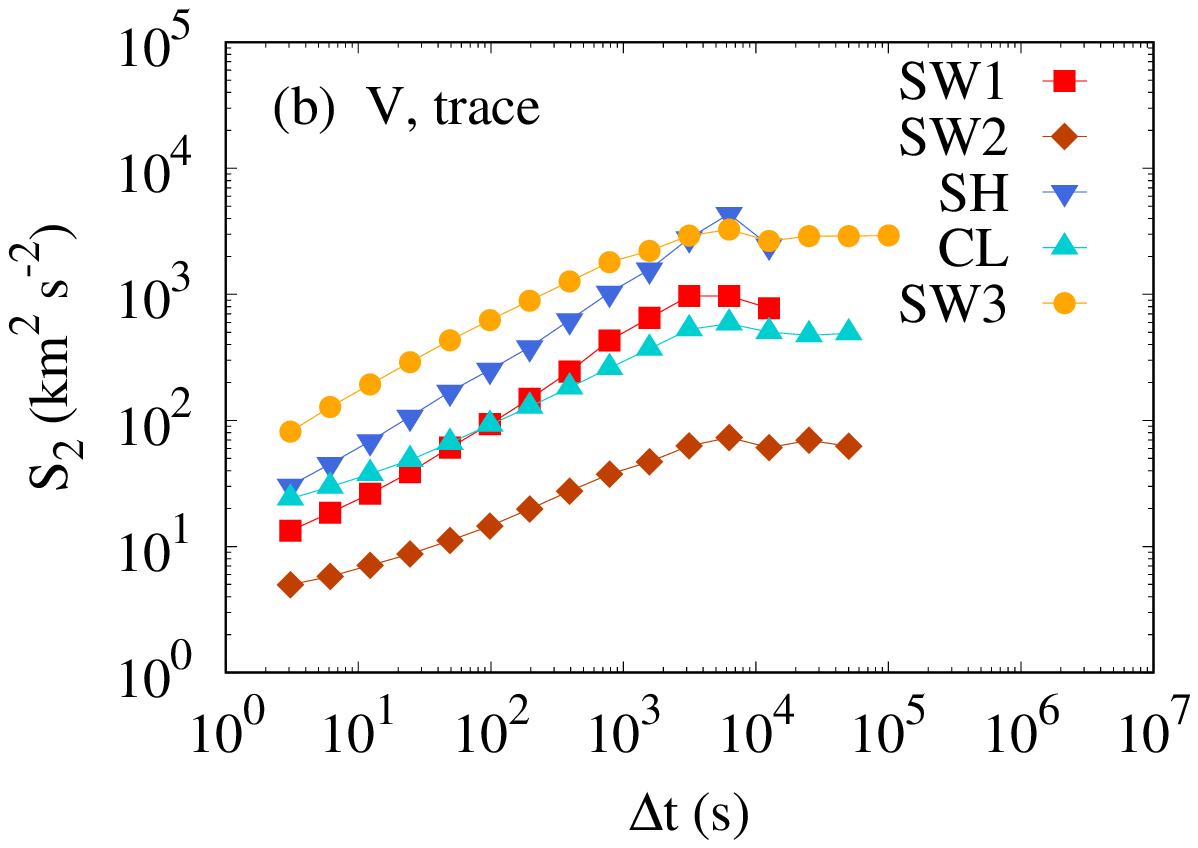}\includegraphics[width=0.33\textwidth]{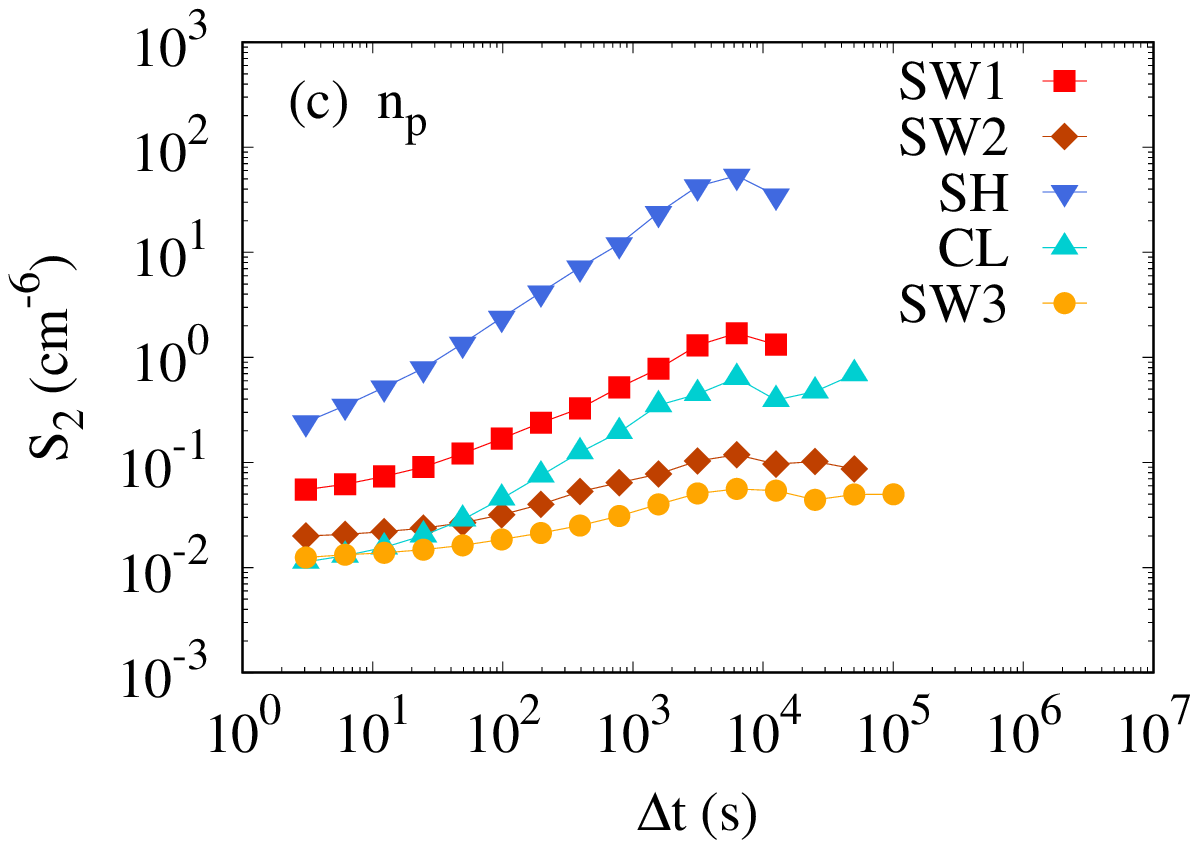}
    \includegraphics[width=0.33\textwidth]{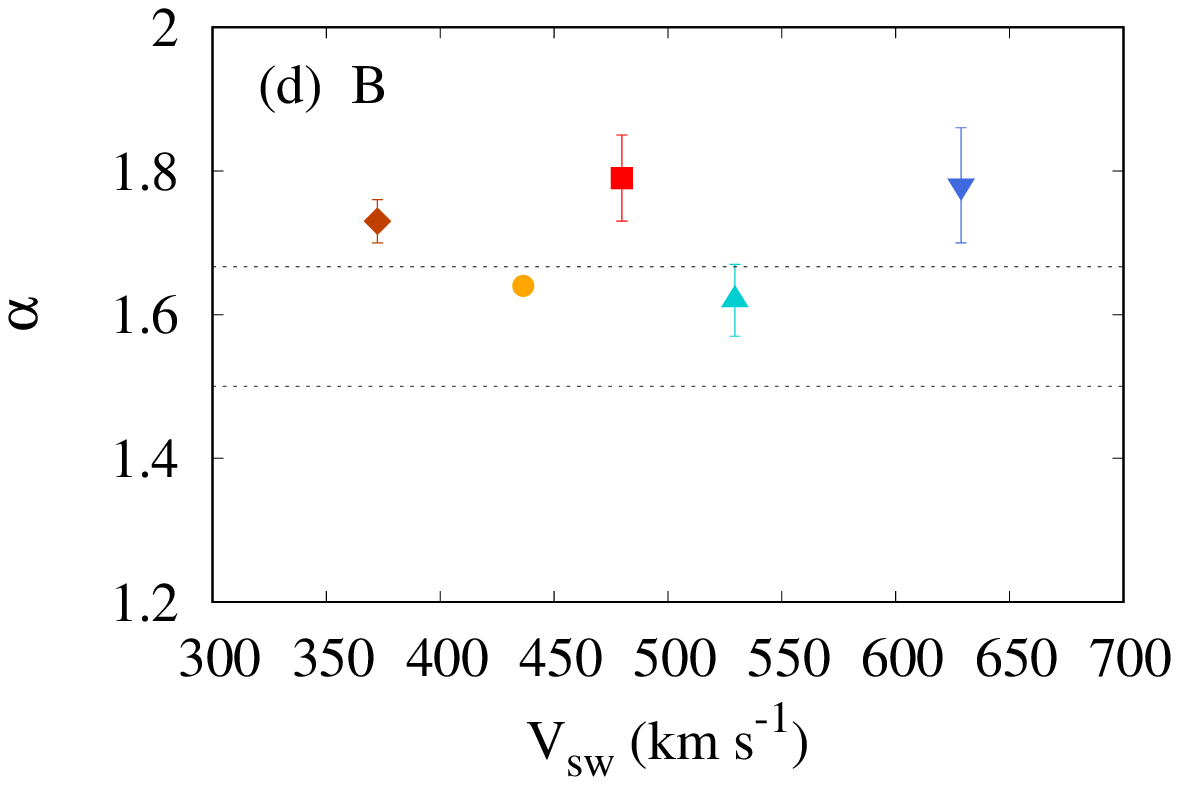}\includegraphics[width=0.33\textwidth]{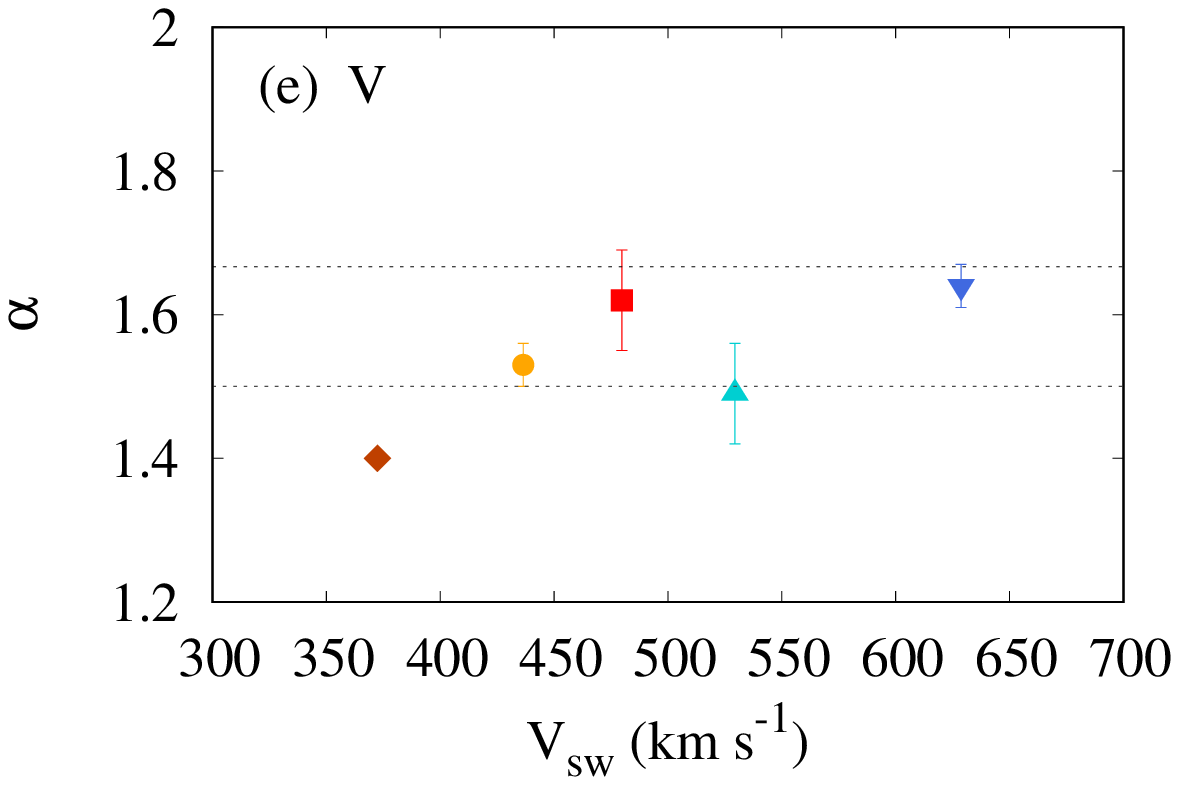}\includegraphics[width=0.33\textwidth]{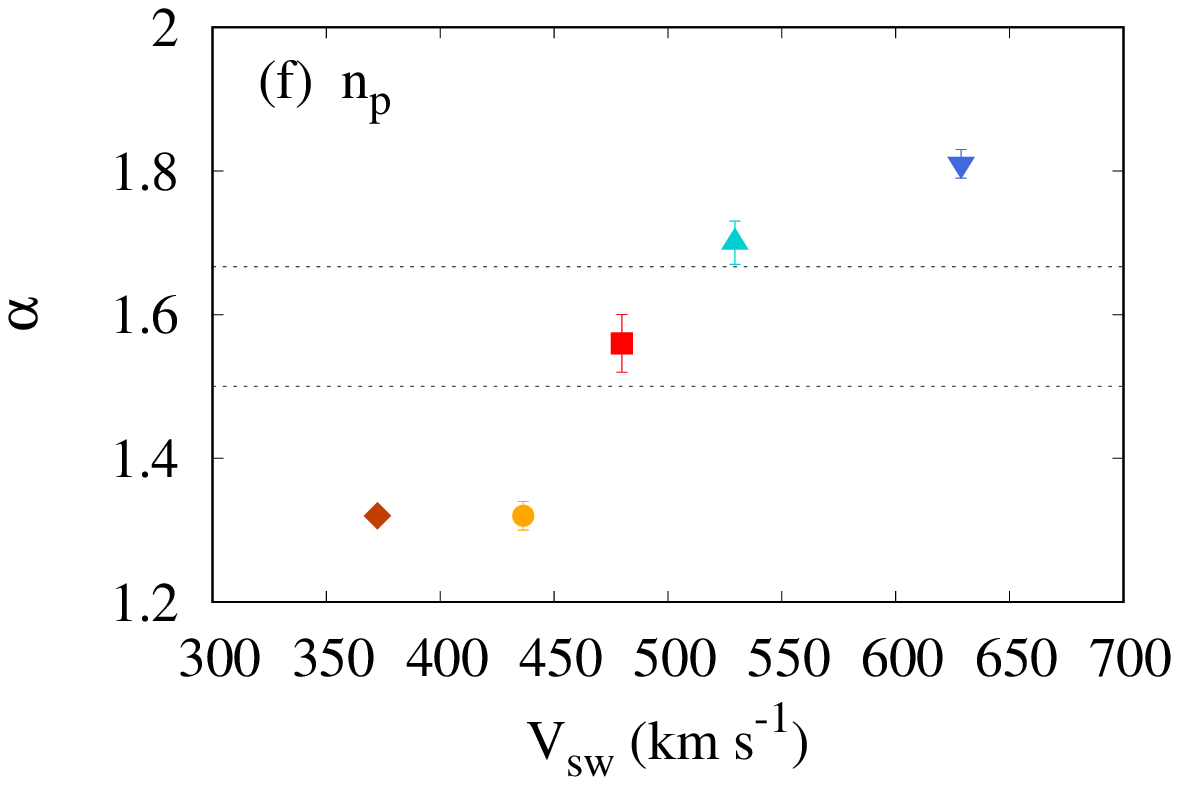}
    \includegraphics[width=0.33\textwidth]{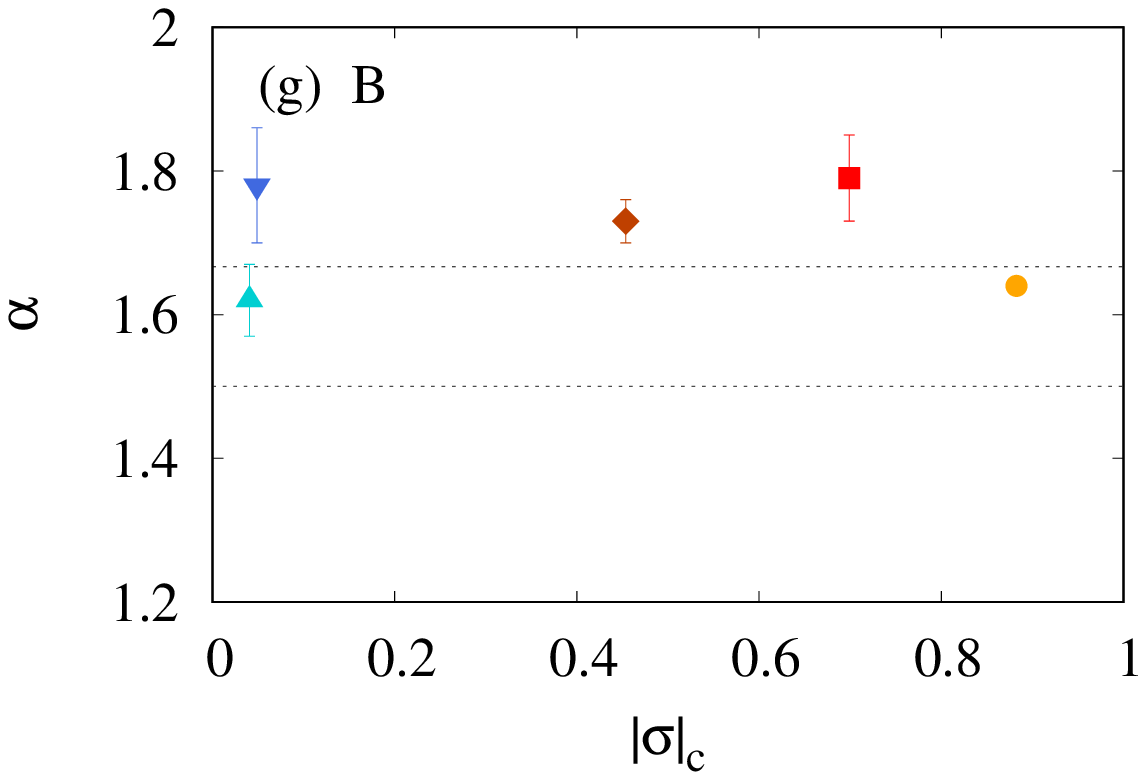}\includegraphics[width=0.33\textwidth]{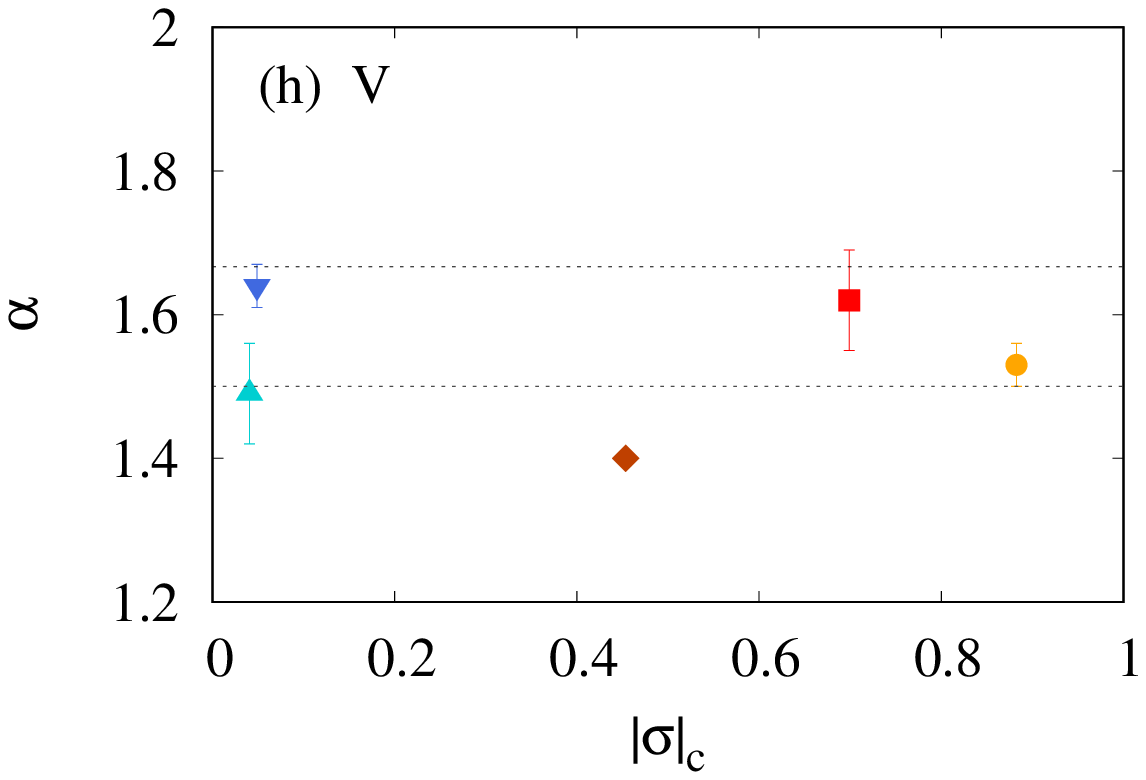}\includegraphics[width=0.33\textwidth]{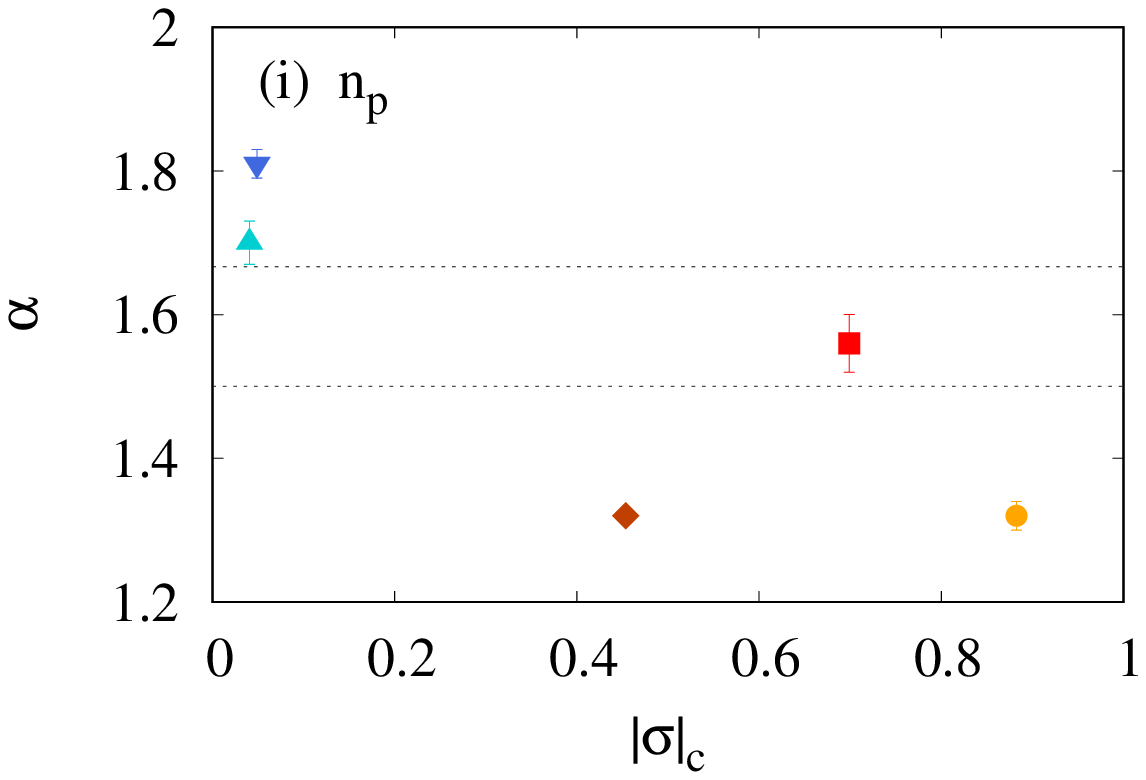}
    \includegraphics[width=0.33\textwidth]{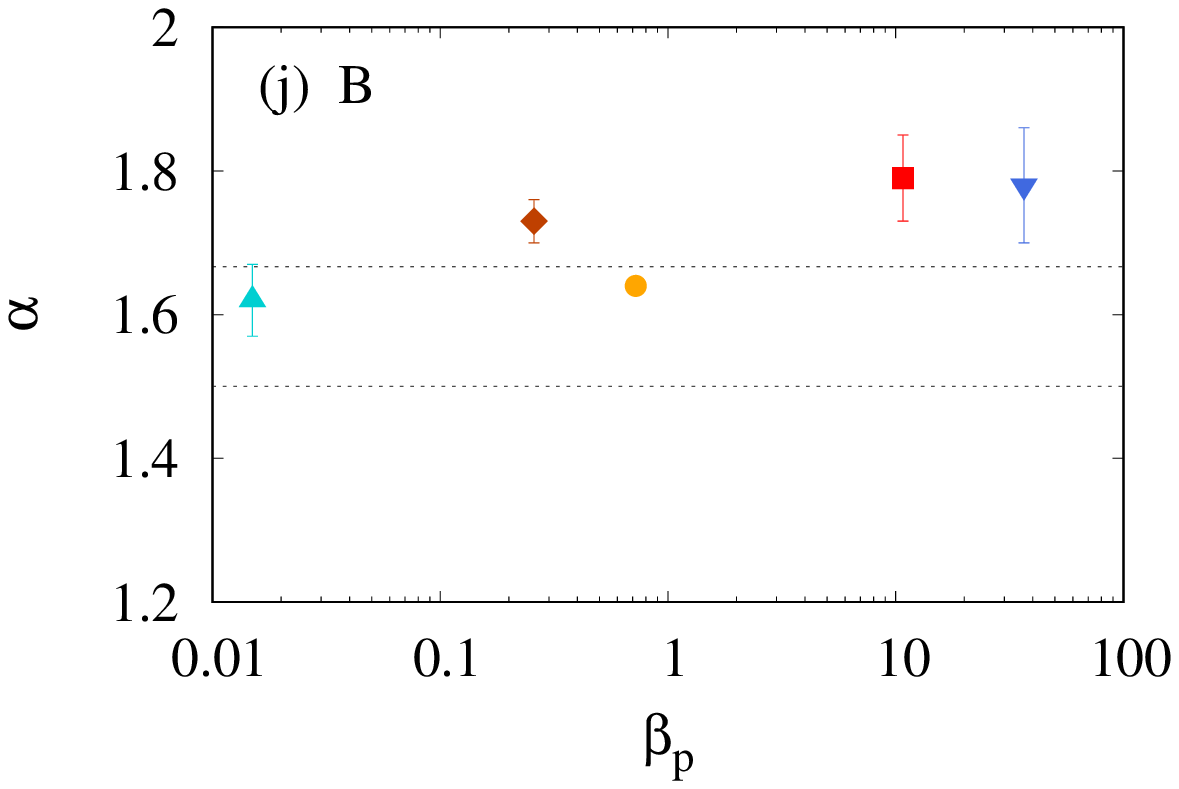}\includegraphics[width=0.33\textwidth]{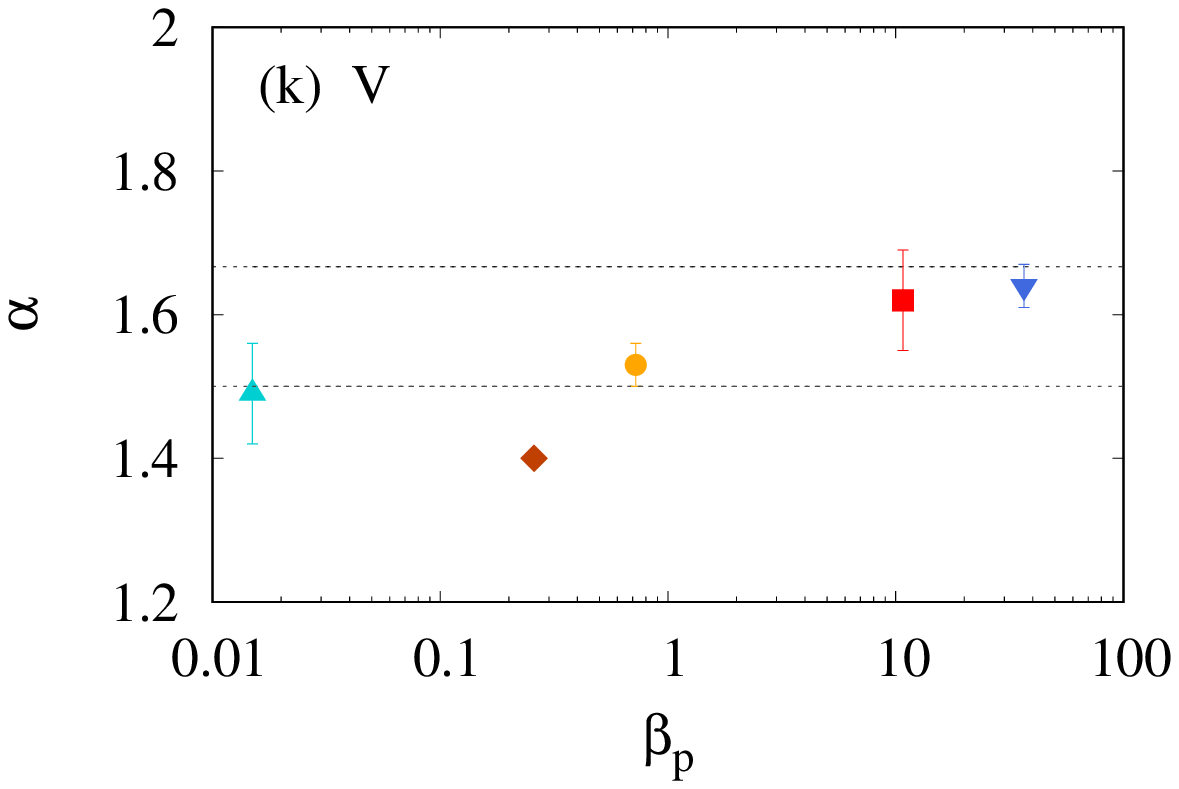}\includegraphics[width=0.33\textwidth]{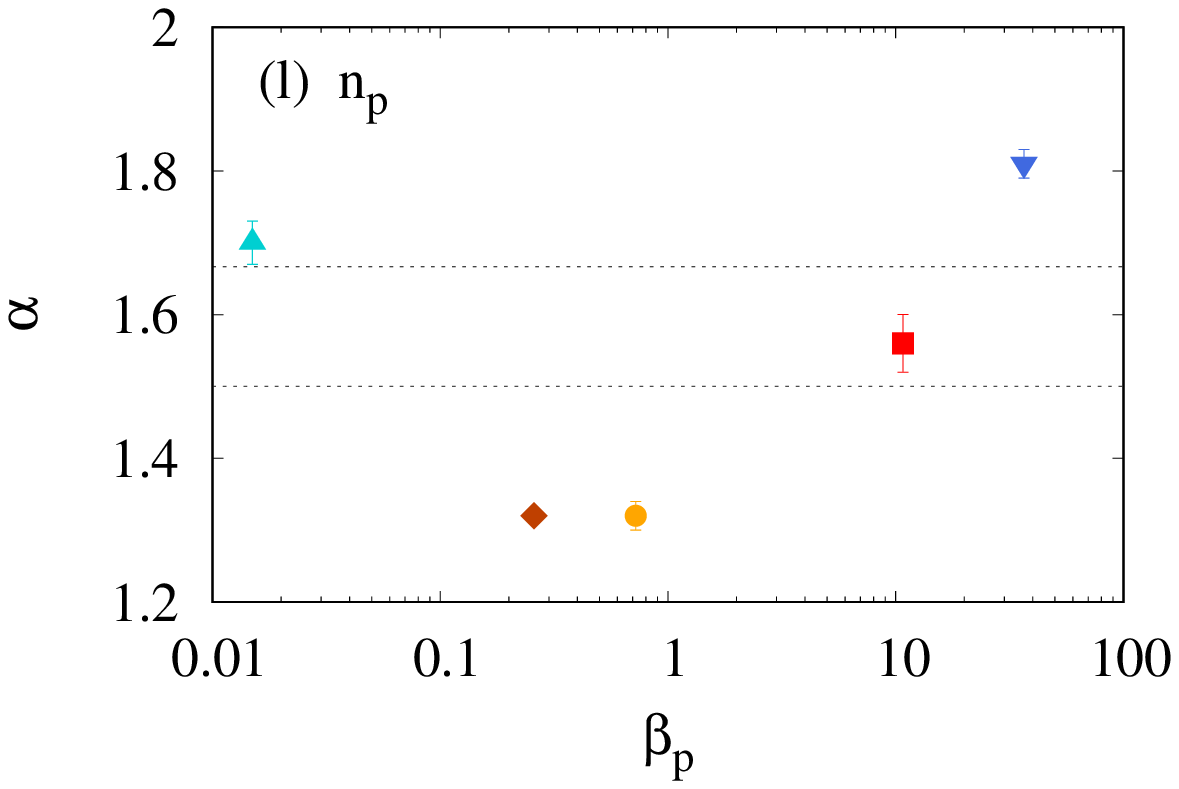}
    \caption{Panels $a$---$c$: second-order structure functions $S_2$ for magnetic field (panel $a$) and velocity (panel $b$) fluctuations, averaged over the three components and labeled as ``trace'', and for density fluctuations (panel $c$), for all five regions. Different quantities have different units.
   Panels $d$---$f$: spectral index $\alpha$ versus the solar wind speed $V_{sw}$. For velocity and magnetic field, values are averaged over the three field components, error bars being the standard deviation. For the density, error bars are the uncertainty of the power-law fit. 
   Panels $g$---$i$ and $j$---$l$: same as central panels but versus the normalized cross-helicity $|\sigma_c|$ and plasma $\beta_p$. 
   All symbols are color-coded for different intervals as described by the labels in panels $a$---$c$, according to Figure~\ref{fig:data}. The dotted horizontal lines indicate the standard values 5/3 (black) and 3/2 (gray).}
    \label{fig:turbo}
\end{figure}

The spectral properties of the magnetic, velocity and density fluctuations were estimated using the second-order structure function $S_2(\Delta t)$, and are plotted in panels $a$, $b$ and $c$ of Figure~\ref{fig:turbo}, respectively. For velocity and magnetic field, the sum over the three components was performed to obtain the equivalent of the spectrum trace. 
The CME sheath region has much larger power for magnetic and density fluctuations, in agreement with previous observations~\citep{Kilpua2020,Pitna2021}, while this is not the case for the velocity.
The structure functions of all fields show power-law scaling for all samples, in a range of scales corresponding to the typical inertial range of turbulence in the solar wind at 1 AU (panels $a$---$c$ of Figure~\ref{fig:turbo}). This extends from $\approx$10 s, close to the estimated ion-cyclotron timescale, to $\approx$1 h, of the order of the typical correlation scale. 
Nonlinear fits to a power-law $S_2(\Delta t)\propto \Delta t^{\alpha-1}$, with $-\alpha$ the spectral exponent~\citep{Frisch1995}, were performed in such range (not shown).
The spectral exponents so estimated are plotted in panels $d$---$i$ of Figure~\ref{fig:turbo} as functions of solar wind parameters. 
In the case of velocity and magnetic field, the exponents have been averaged over the three field components, the error bars indicating the standard deviation. 
As typically observed in solar wind plasmas~\citep{BrunoCarbone2013}, for all cases the exponents $\alpha$ are in the range $1.3$---$1.85$, roughly compatible with the presence of a turbulent cascade. 
The magnetic field components display the standard Kolmogorov spectral exponents, close to 5/3, which do not seem to depend on the flow speed $V_{sw}$ (panel $d$) or on the unsigned normalized cross helicity $|\sigma_c|$ (panel $g$), and only weakly depends on $\beta_p$ (panel $j$). 
On the other hand, the velocity has slightly shallower exponents, being closer to the value for strongly Alfv\'enic turbulence 3/2. In this case, the exponent is smaller for slower solar wind (panel $e$)~\citep{Pitna2021} smaller $\beta_p$ (panel $k$), while there is no clear dependence on the cross-helicity (panel $h$). 
Finally, the density exponents show clear dependence of the solar wind speed (panel $f$) $\beta_p$ (panel $k$).
They are closer to 5/3 for the faster CME plasma, where cross-helicity is nearly zero, and closer or smaller than 3/2 in the weakly compressible preceding and trailing solar wind, where the cross-helicity is larger (panel $i$). 
Therefore, while velocity and magnetic field components seem to have similar spectra inside or outside of the CME, the proton density shows substantial differences between the two cases, possibly due to the more compressive nature of the CME plasma~\citep{Kilpua2020,Pitna2021}.

\begin{figure}
    \centering
    \includegraphics[width=0.33\textwidth]{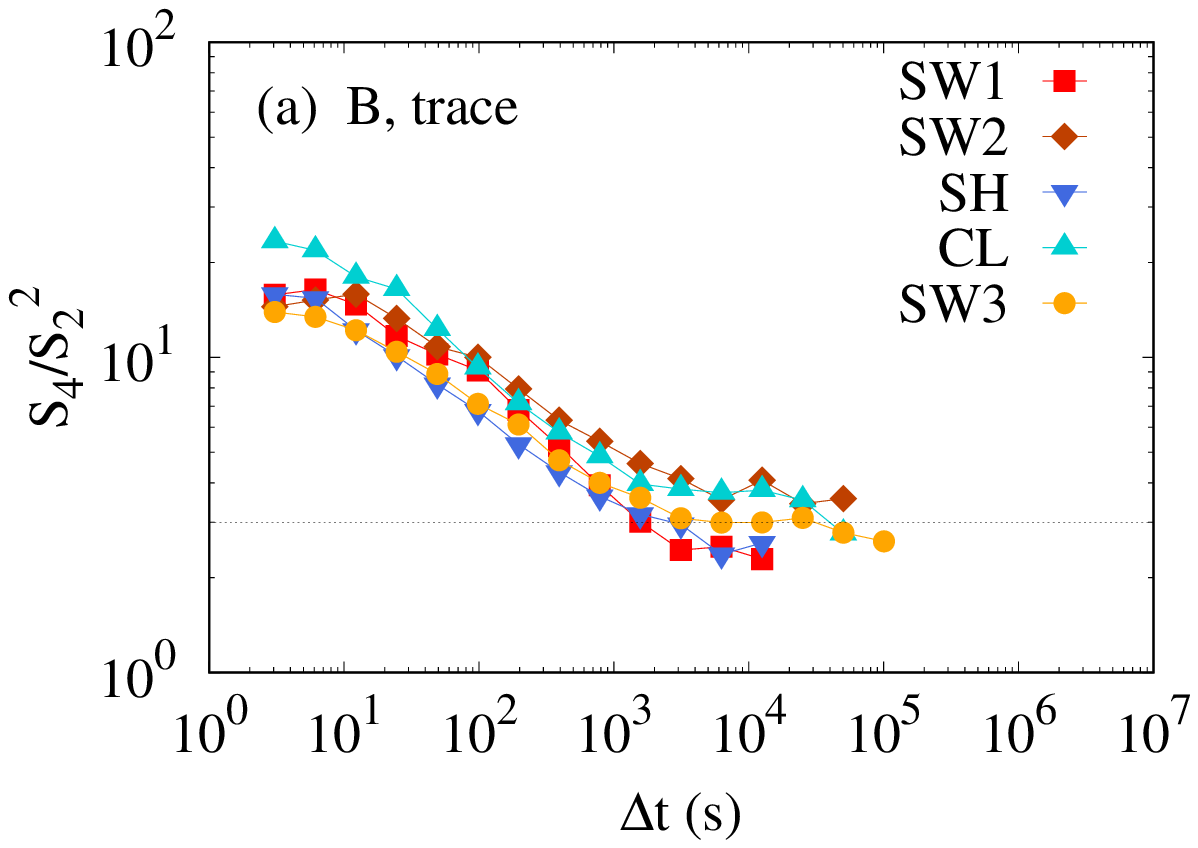}\includegraphics[width=0.33\textwidth]{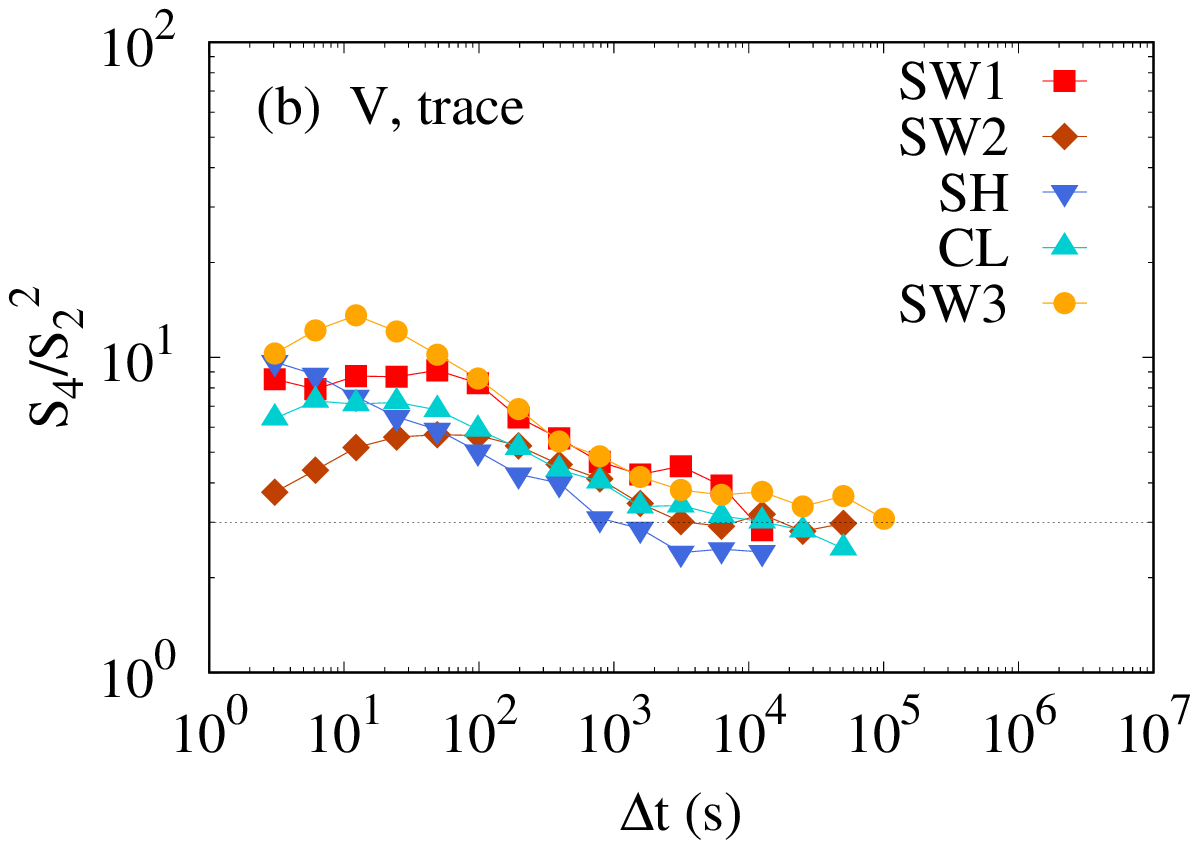}\includegraphics[width=0.33\textwidth]{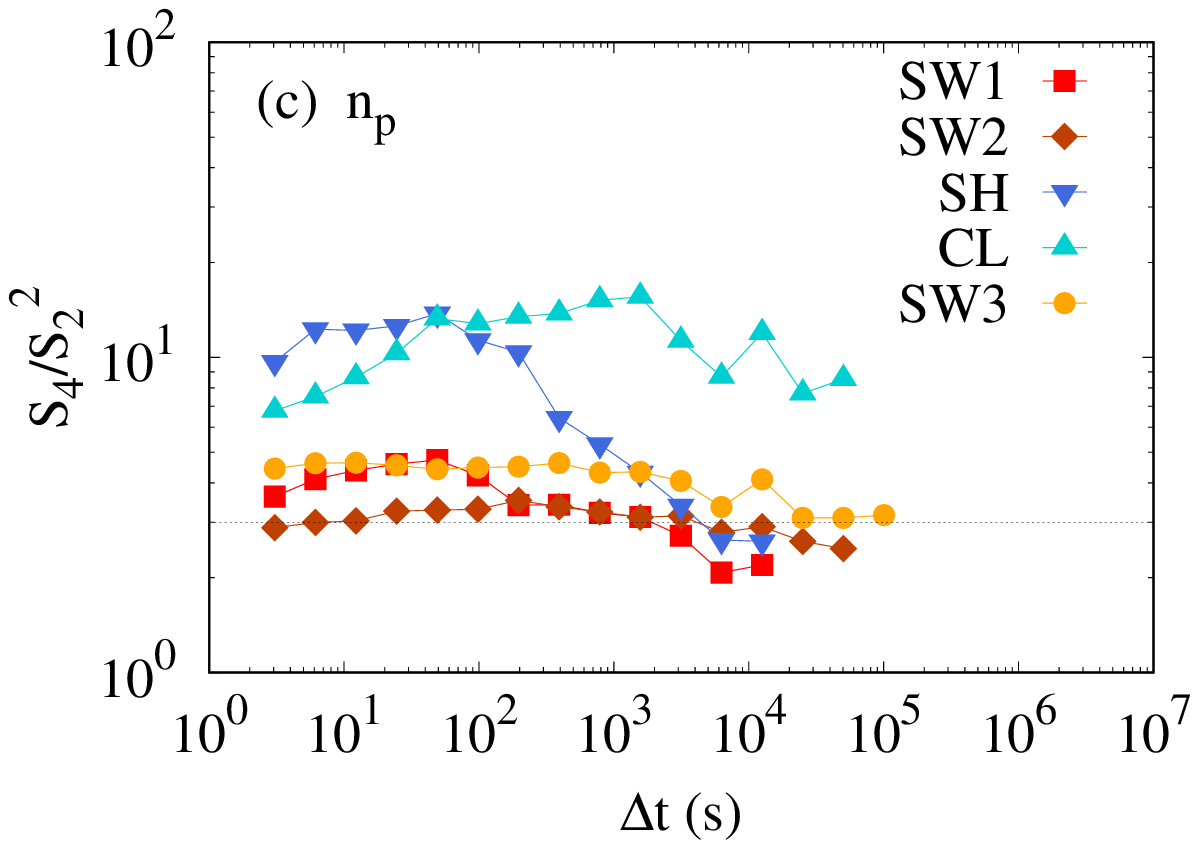}
    \includegraphics[width=0.33\textwidth]{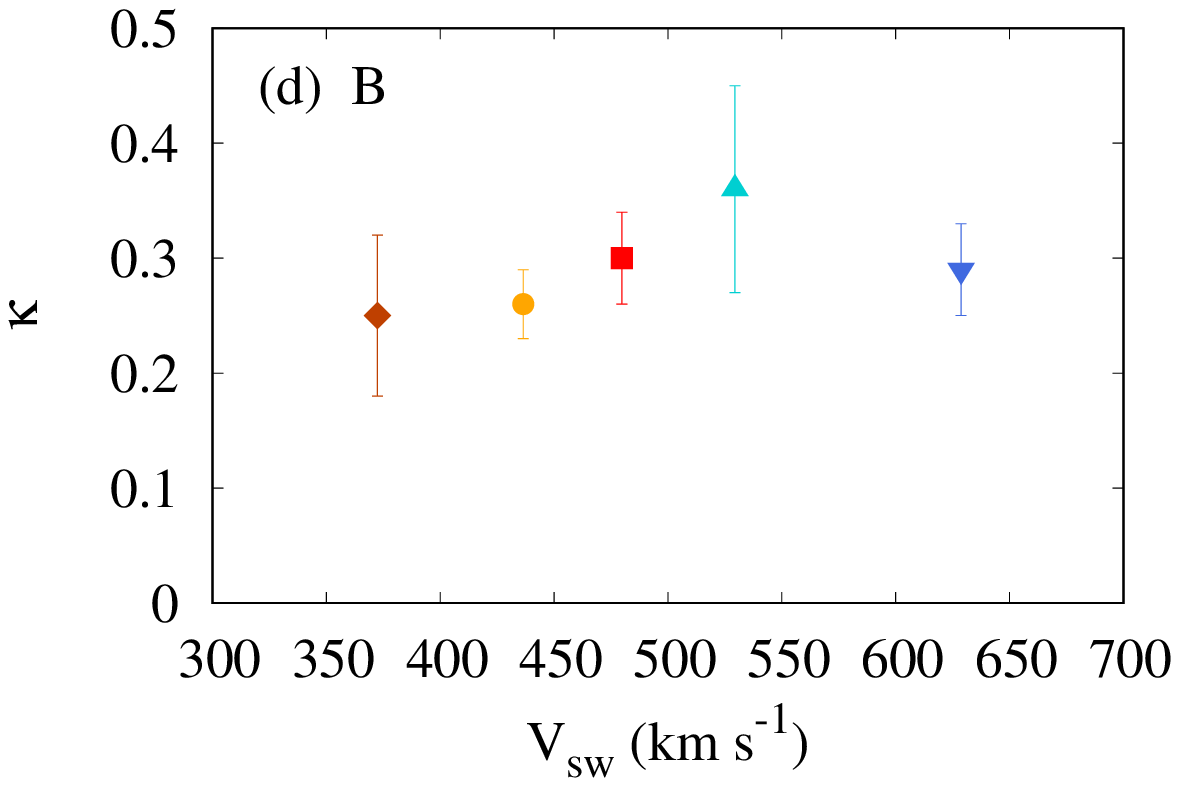}\includegraphics[width=0.33\textwidth]{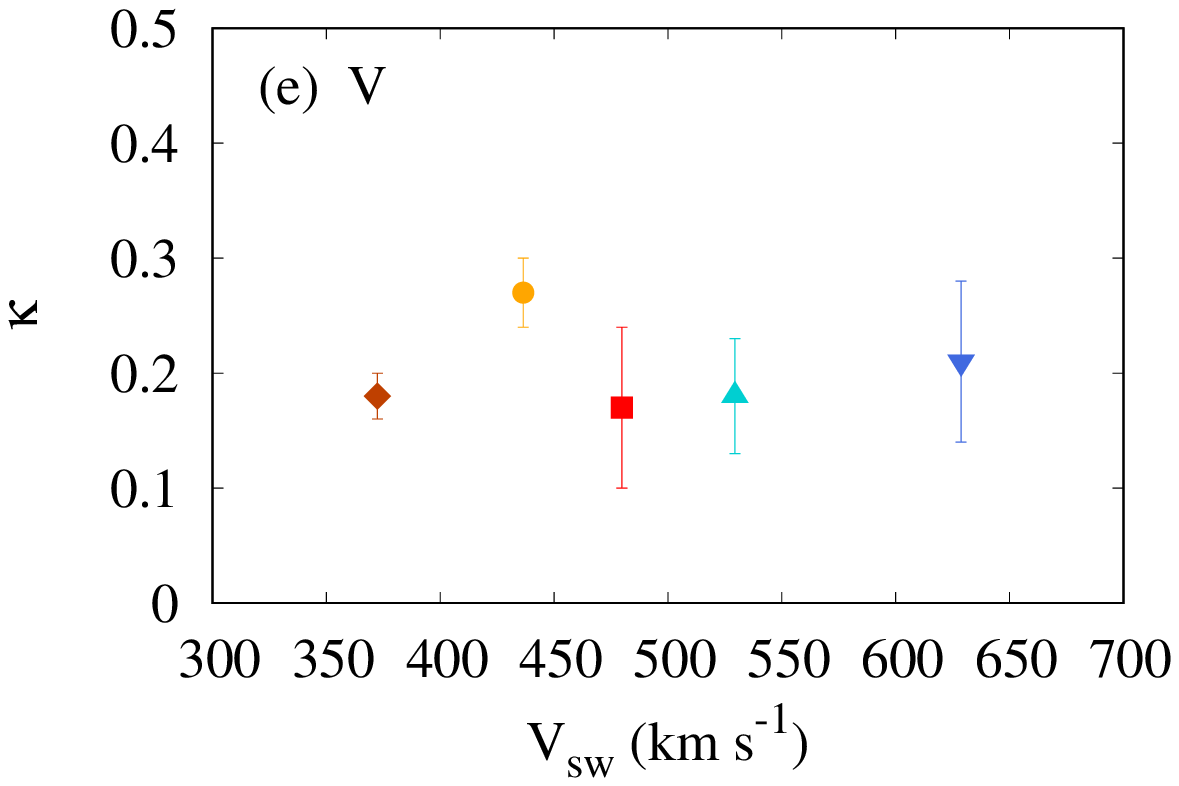}\includegraphics[width=0.33\textwidth]{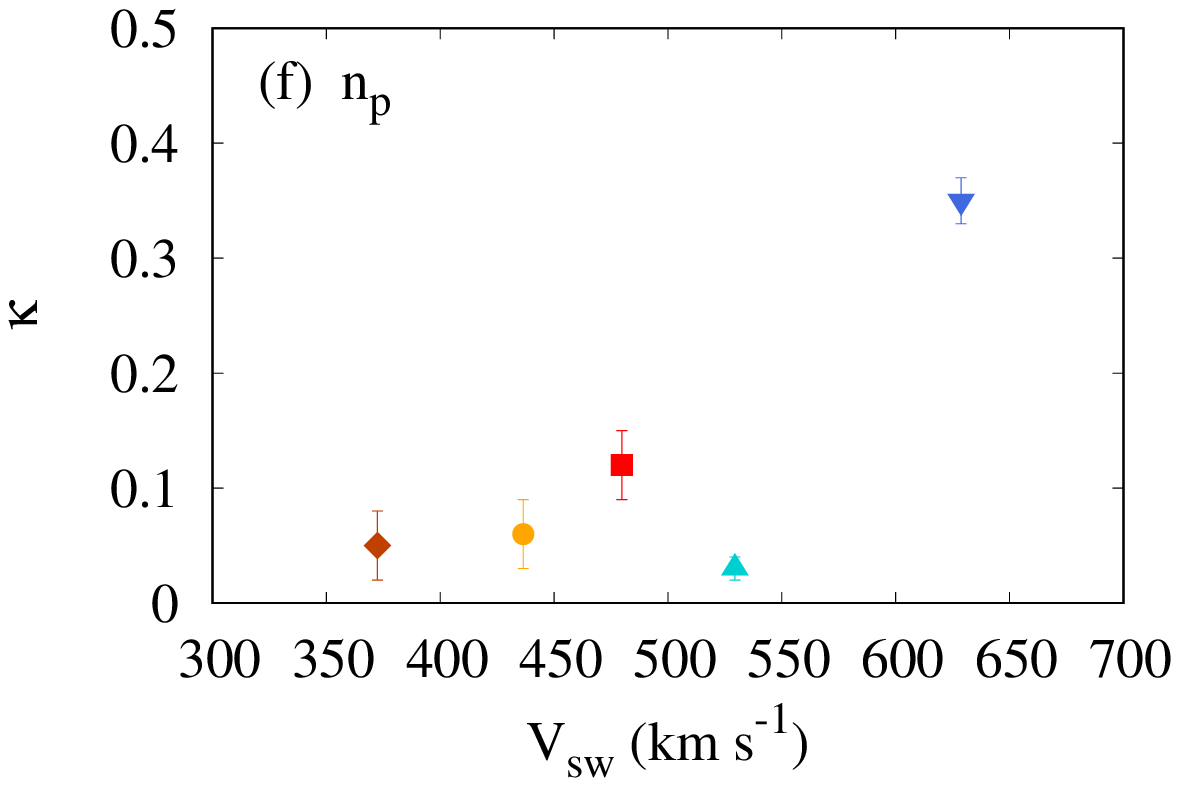}
    \includegraphics[width=0.33\textwidth]{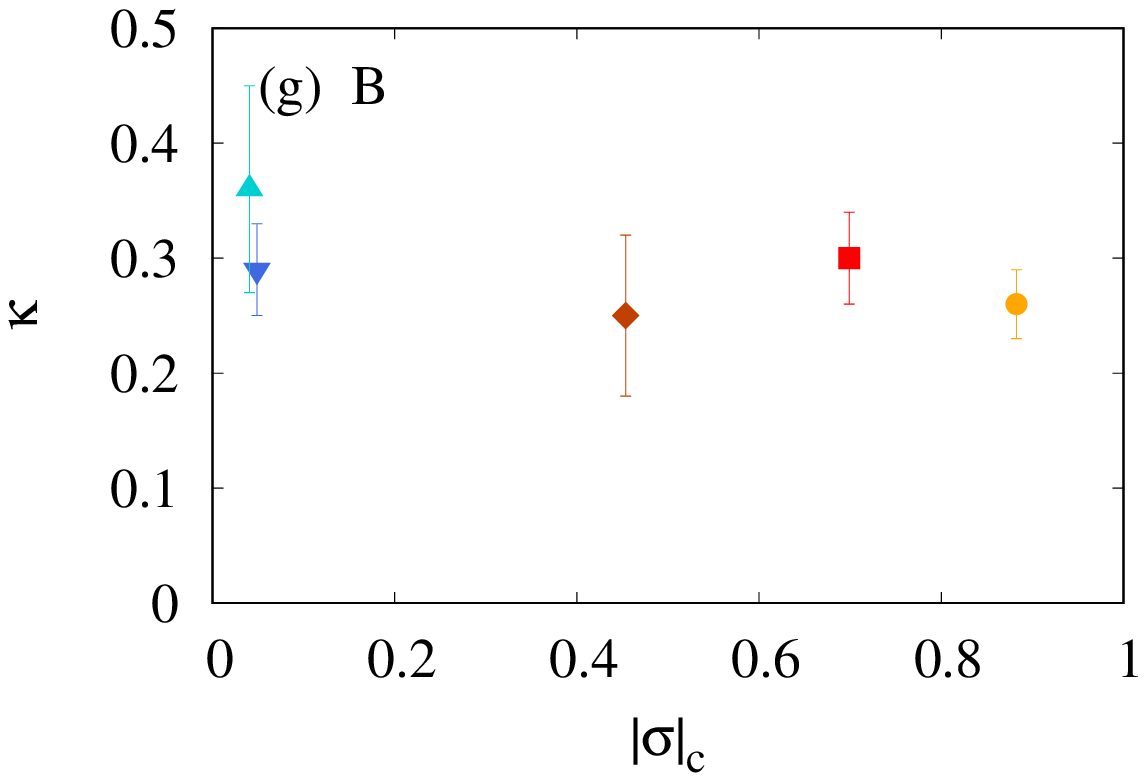}\includegraphics[width=0.33\textwidth]{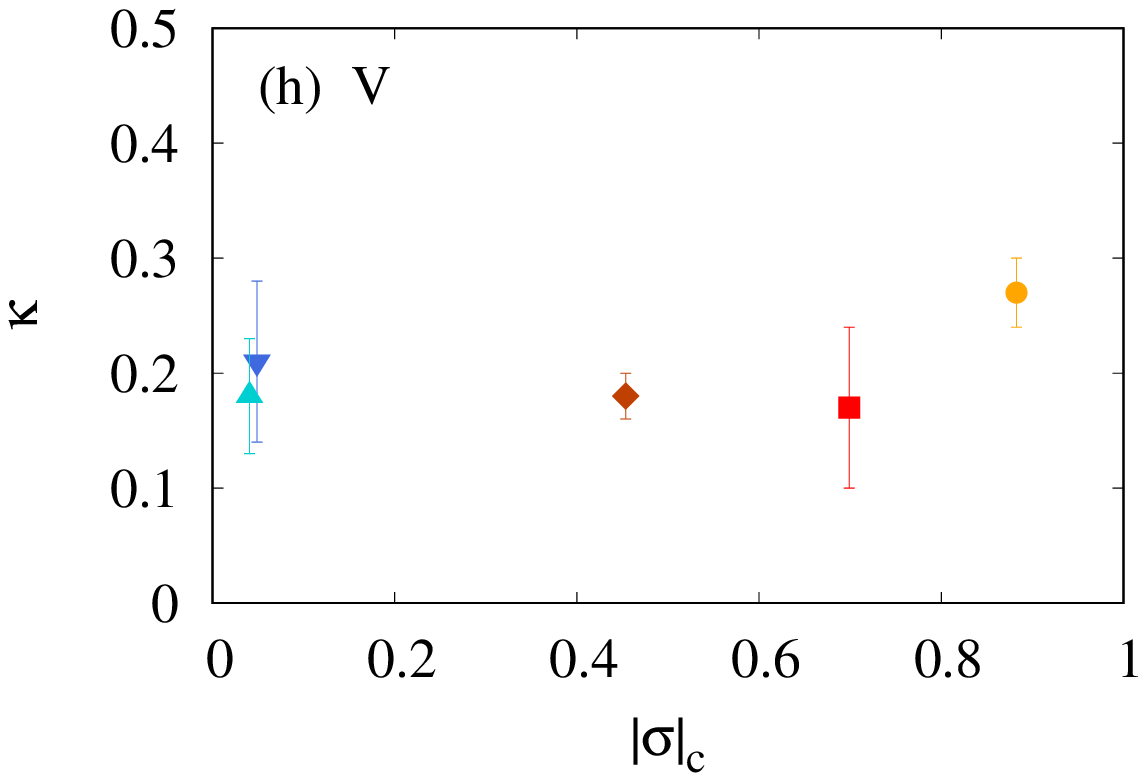}\includegraphics[width=0.33\textwidth]{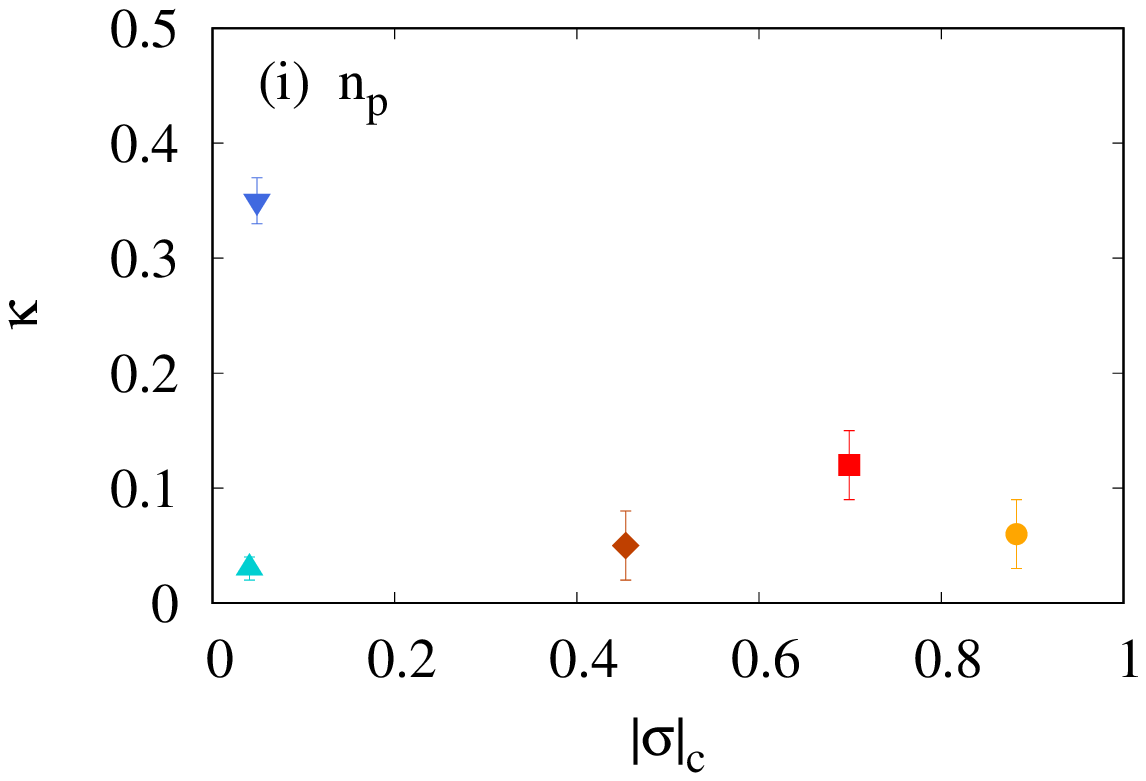}
    \includegraphics[width=0.33\textwidth]{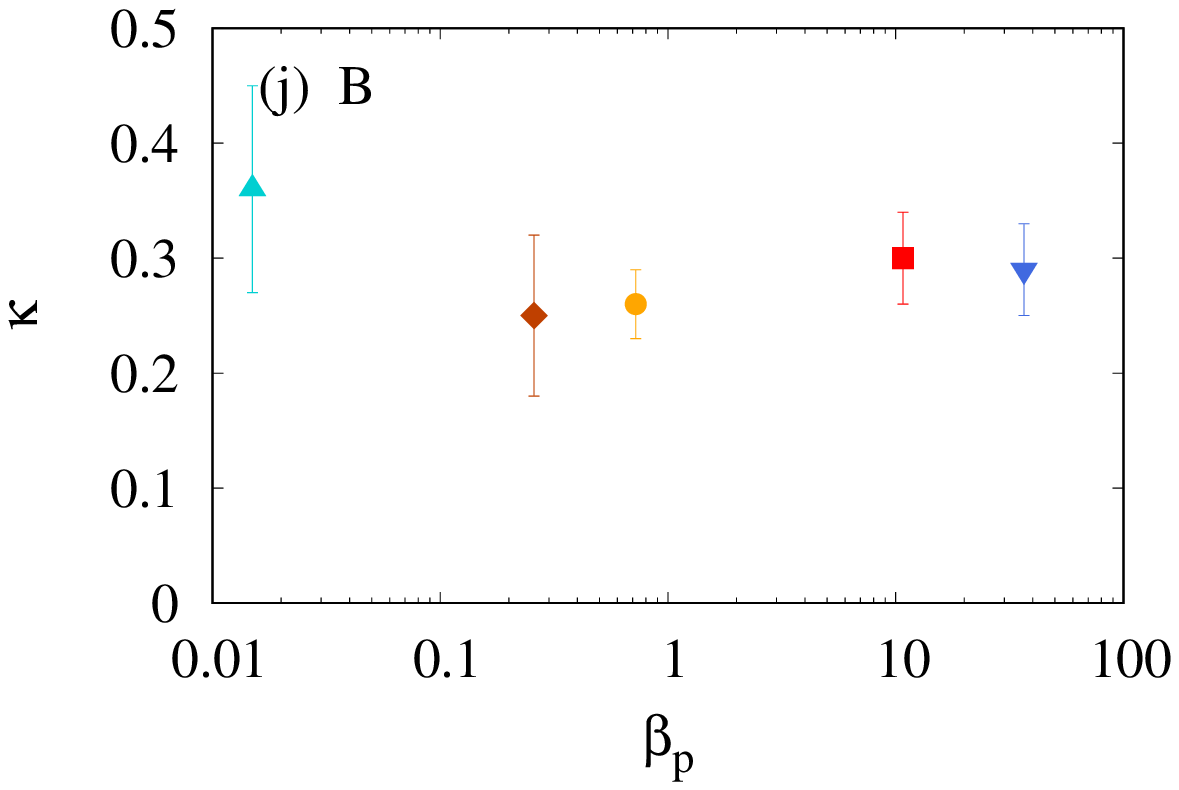}\includegraphics[width=0.33\textwidth]{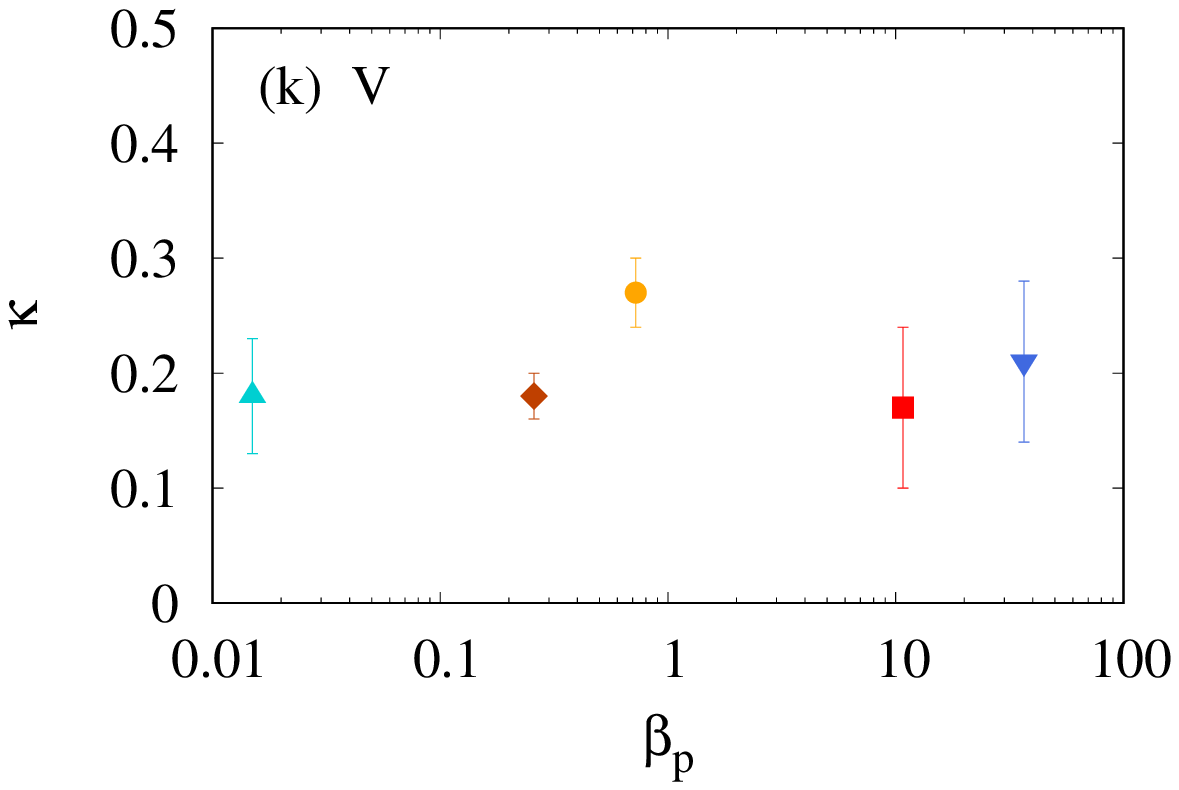}\includegraphics[width=0.33\textwidth]{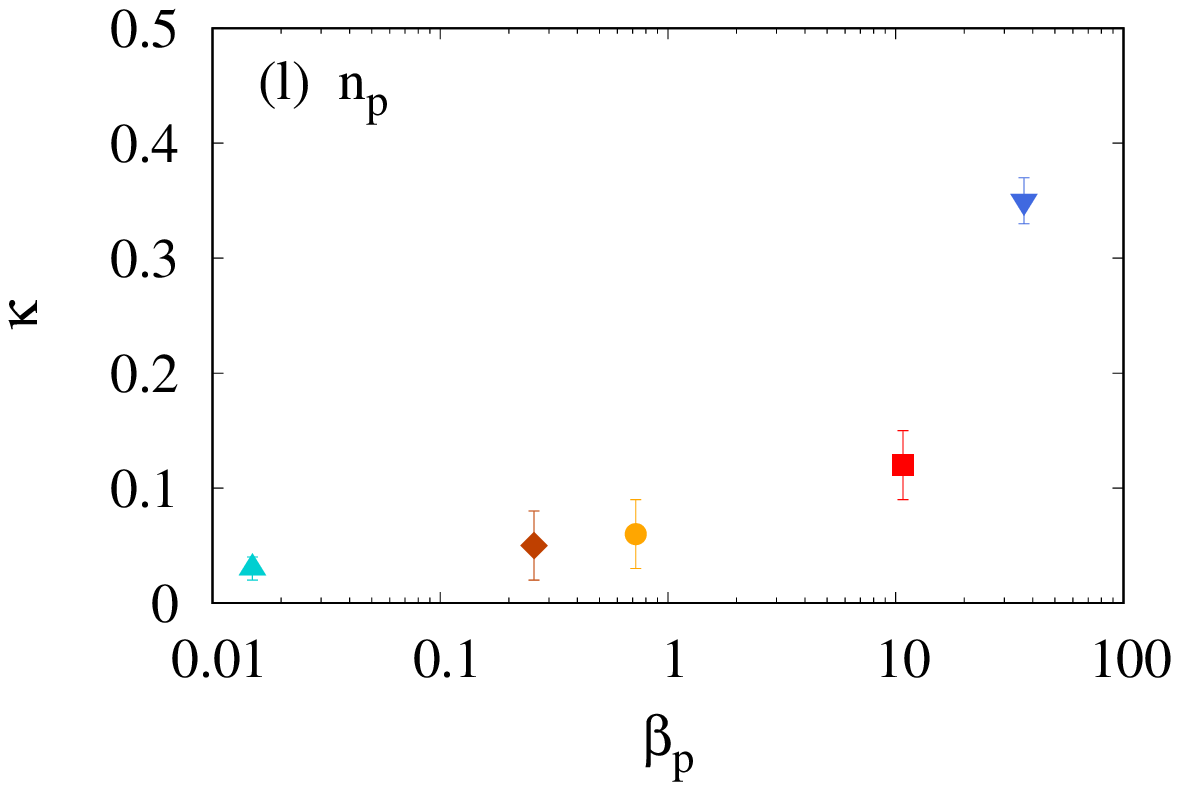}
    \caption{Panels $a$---$c$: the flatness $F=S_4/S_2^2$ for magnetic field (panel $a$) and velocity (panel $b$) fluctuations, averaged over the three components and labeled as ``trace'', and for density fluctuations (panel $c$), for all five regions. The gray horizontal line indicate the Gaussian value $F=3$.
    Panels $d$---$f$: flatness scaling exponent $\kappa$ versus solar wind speed $V_{sw}$. For velocity and magnetic field, values are averaged over the field components, error bars being the standard deviation. For the density, error bars are the uncertainty of the power-law fit.
    Panels $g$---$i$: same as central panels but versus the normalized cross-helicity $|\sigma_c|$ and plasma $\beta_p$. 
    All symbols are color-coded for different intervals as described by the labels in panels $a$---$c$, according to Figure~\ref{fig:data}.}
    \label{fig:flatness}
\end{figure}

For assessing the intermittency properties of the turbulence, the flatness $F(\Delta t)= S_4(\Delta t)/S^2_2(\Delta t)$ was evaluated for each selected interval~\citep{DudokDeWit2013}. 
The flatness describes the scale-dependent deviation from Gaussian ($F=3$) of the field increment distributions, quantitatively accounting for the presence of high tails arising from the generation of small-scale intense, intermittent structures.
The scaling properties of the flatness, related to the level of intermittency of the time series, are visible in Figure~\ref{fig:flatness}, where again for velocity and magnetic field the values have been averaged over the components.
The presence of a power-law $F(\Delta t)\propto \Delta t^{-\kappa}$ is a consequence of the scale invariance of the magnetohydrodynamic (MHD) equations in the turbulence inertial range and of the anomalous scaling of the structure functions~\citep{Frisch1995,BrunoCarbone2013}, and supports the existence of a turbulent cascade. 
The scaling exponent $\kappa$ is related to the efficiency of the energy transfer process or, equivalently, to the fractal co-dimension of the most intermittent structures~\citep{Castaing1990,Carbone2014}. 
Larger $\kappa$ corresponds to more rapid generation of small-scale structures, or less space-filling structures of smaller fractal dimension. 
A power-law fit was performed for all fields and components and for all samples. 
The flatness analysis reveals that intermittency, as measured through the flatness values and the scaling exponent $\kappa$, is present in all samples and for all components of velocity and magnetic fields, the observed parameters being in agreement with previous observations~\citep[see e.g.][and references therein]{Quijia2021}. 
Conversely, the proton density has very low to no intermittency in the solar wind intervals, while in the CME sheath region it is comparable with the other fields~\citep{Kilpua2020,Carbone2021}. In the anomalous CME cloud the flatness is large at all scales. 
Looking at panels $d$---$f$ of Figure~\ref{fig:flatness}, the flatness scaling exponent roughly increases with the solar wind speed for magnetic field and proton density, while no clear trend is visible for the velocity.
No dependence on the normalized cross-helicity is observed (panels $g$---$i$). The dependence on $\beta_p$ is negligible for velocity and magnetic field (panels $j$ and $k$) but very clear for the density (panel $l)$.

%
%

%
%


\section{Third-order moment scaling law and energy transfer rate} 
\label{sec:yaglom}

In order to determine the properties of the turbulent cascade, the validity of the Politano-Pouquet (PP) law~\citep{PolitanoPouquet1998} was tested for each of the intervals. 
The PP law describes the global energy transfer across scales in the turbulent cascade, and relates the structure functions of the relevant fields to the energy transfer rate. 
For a turbulent flow, under the assumptions of statistical homogeneity, stationarity, high Reynolds number, and local isotropy, the incompressible MHD equations lead to linear scaling of the mixed third-order structure functions
\begin{equation}
    Y(\Delta t) \equiv \langle \Delta v_L (|\Delta \mathbf{v}|^2+|\Delta \mathbf{b}|^2) - 2 \Delta b_L (\Delta \mathbf{v}\cdot\Delta \mathbf{b})\rangle = 4 \varepsilon V_{sw} \Delta t /3 \, ,
\label{Eq:PP}
\end{equation}
%
where $\varepsilon$ is the mean energy transfer rate of the turbulent cascade and $\Delta \phi_L = \phi_L(t+\Delta t)-\phi_L(t)$ are longitudinal increments of the field component $\phi_L$ in the direction of the solar wind mean flow.
The bulk speed $V_{sw}$ is used for transforming space lags $\ell$ in time lags $\Delta t$ through the Taylor hypothesis, $\ell = V_{sw} \Delta t$~\citep{Taylor1938}. Note that the opposite direction of the increment with respect to the flow reverses the sign of the $r.h.s.$ of equation~\ref{Eq:PP}.
Brackets indicate time average over the sample.
In the solar wind and in near-Earth space, the PP law has been observed in the basic form given above~\citep{MacBride2005,Sorriso-Valvo2007,Smith2009} as well as in more complete forms 
that include Hall effects~\citep{Ferrand,Hellinger,riddhi} and compressibility~\citep{Hadid2018,Andres2019}. 

\begin{figure}
    \centering
    \includegraphics[width=0.33\textwidth]{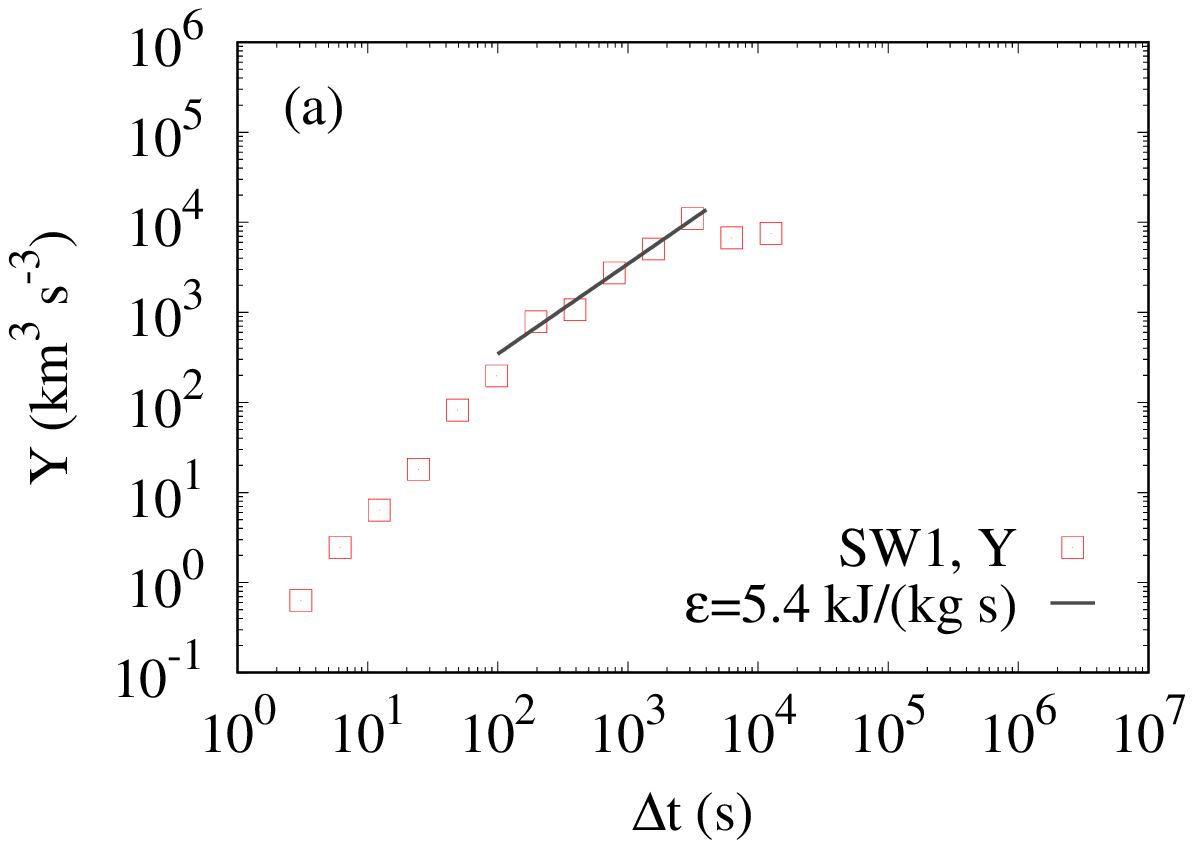}\includegraphics[width=0.33\textwidth]{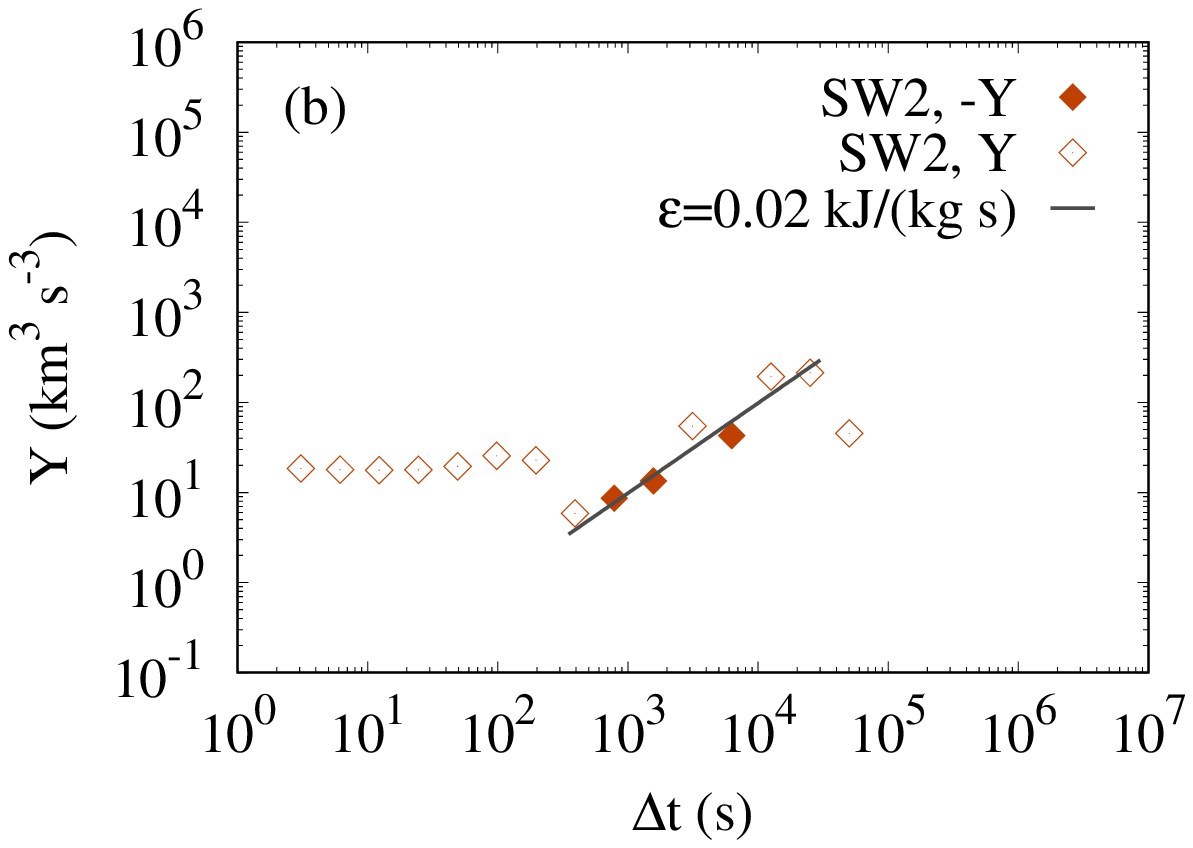}\includegraphics[width=0.33\textwidth]{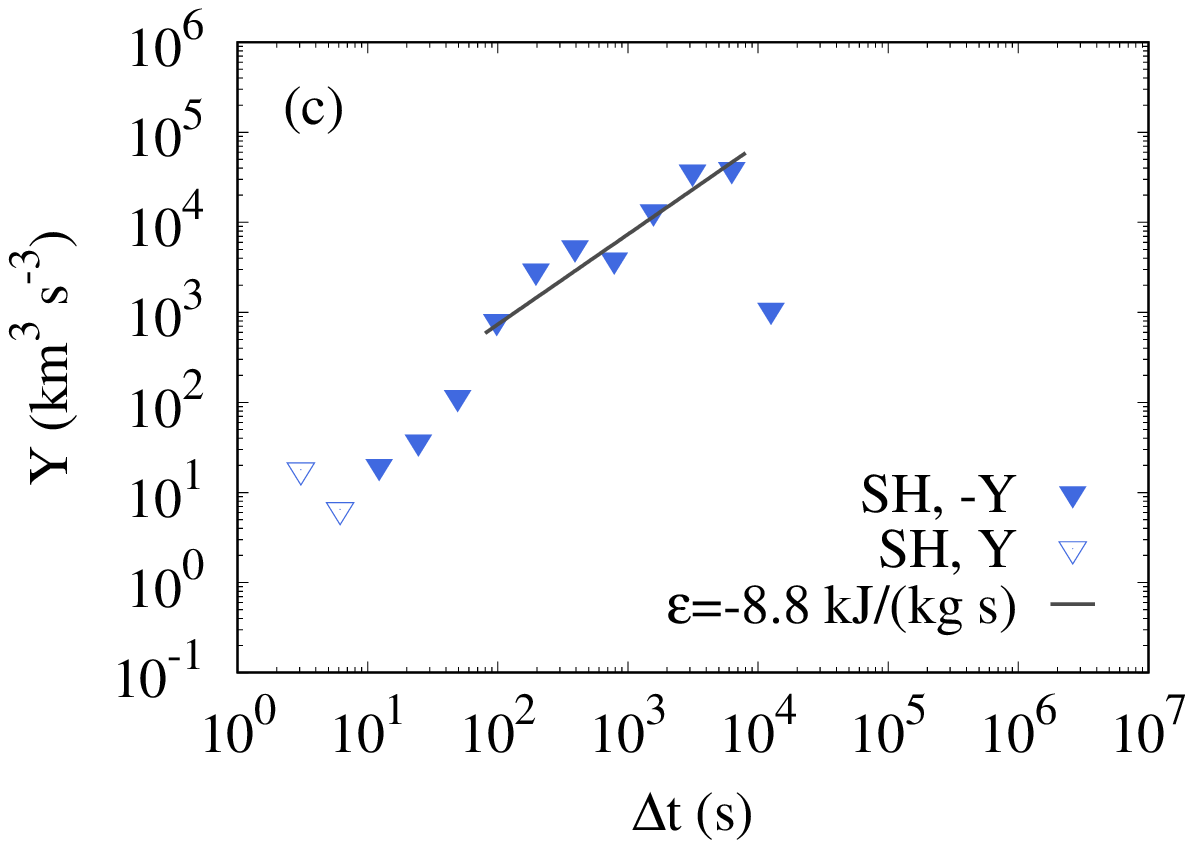}
    \includegraphics[width=0.33\textwidth]{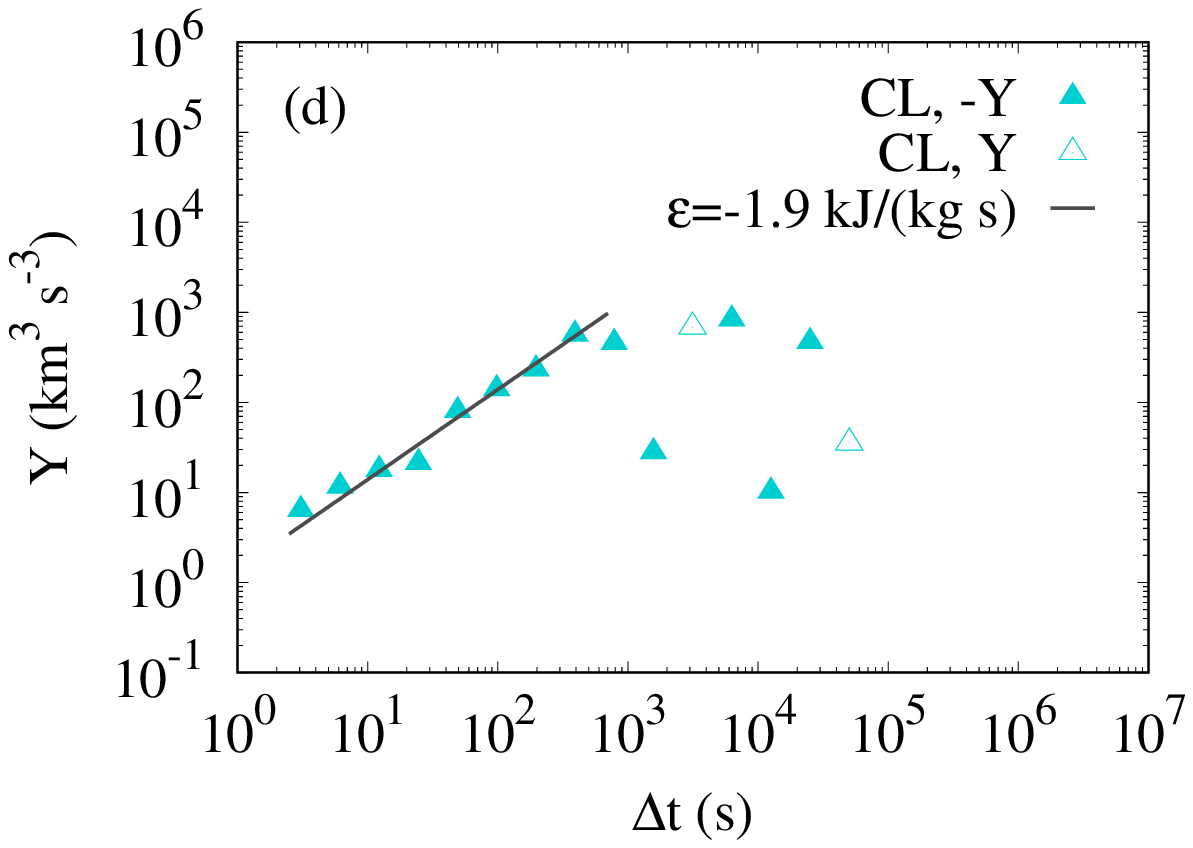}\includegraphics[width=0.33\textwidth]{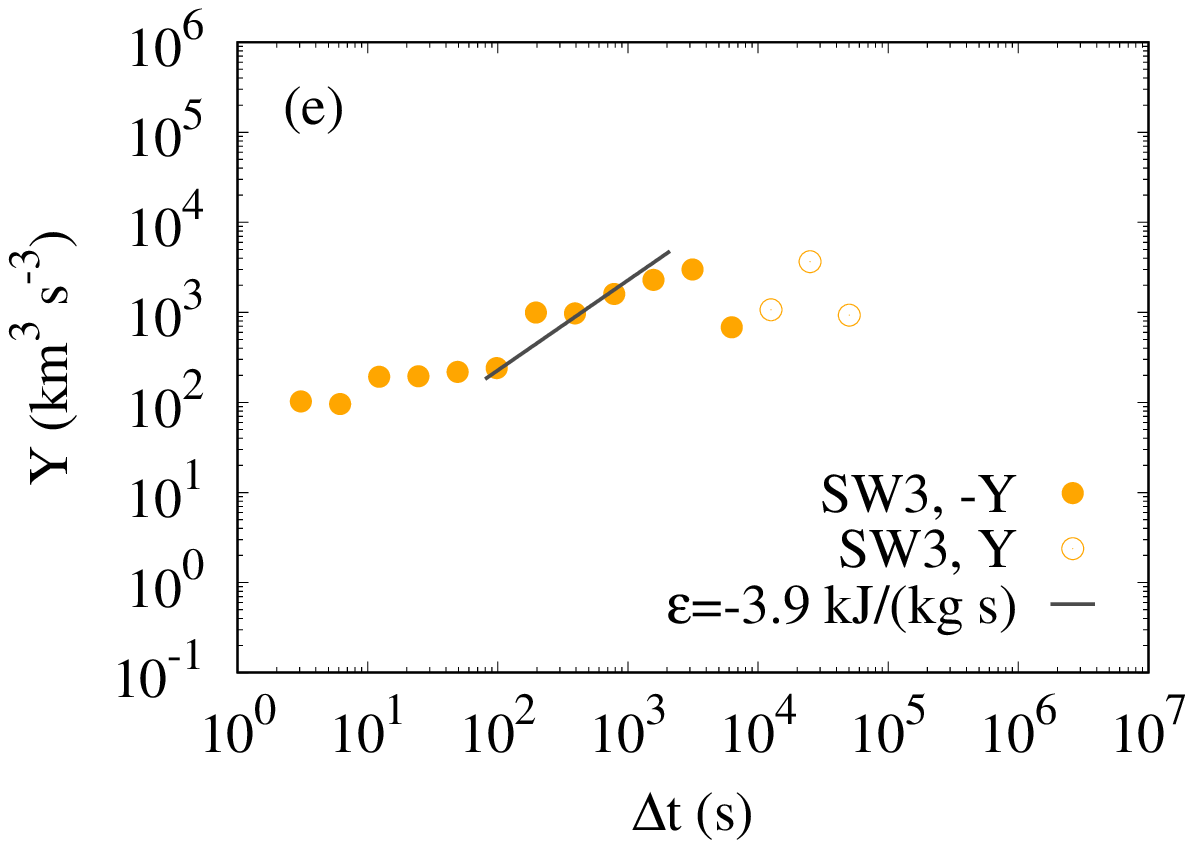}\includegraphics[width=0.33\textwidth]{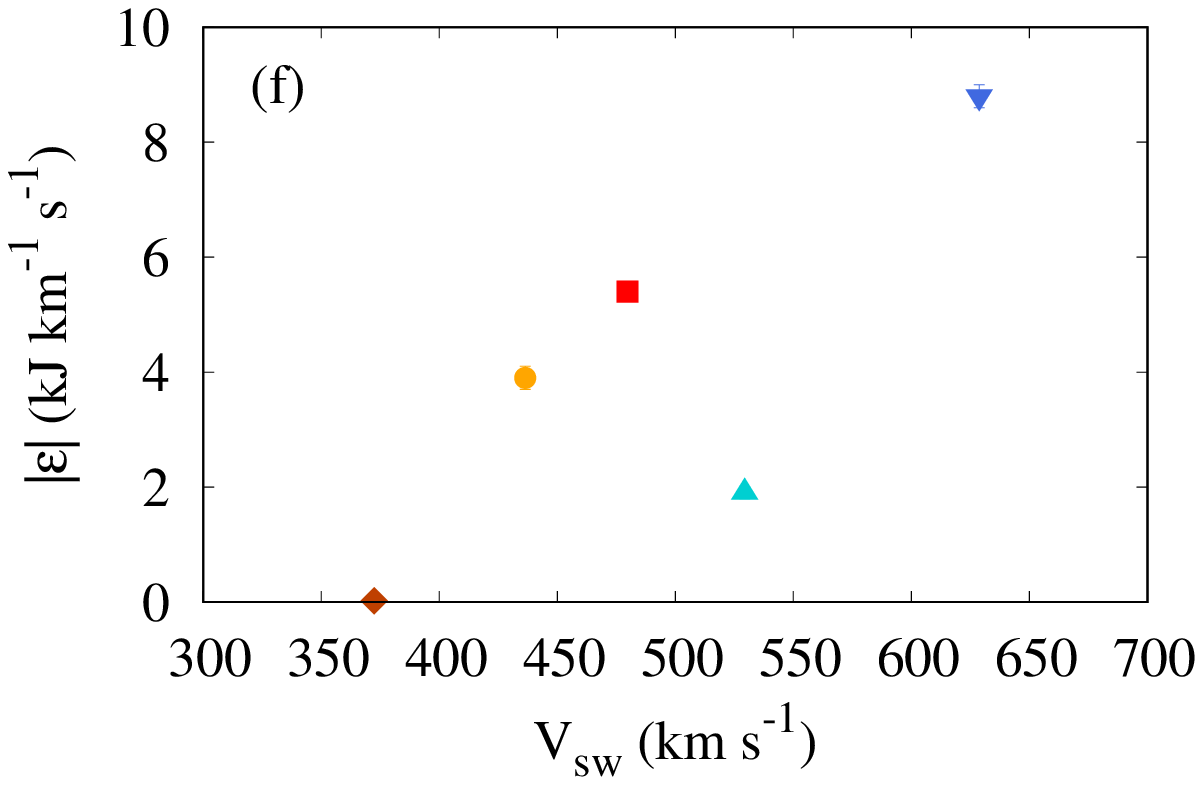}
    \includegraphics[width=0.33\textwidth]{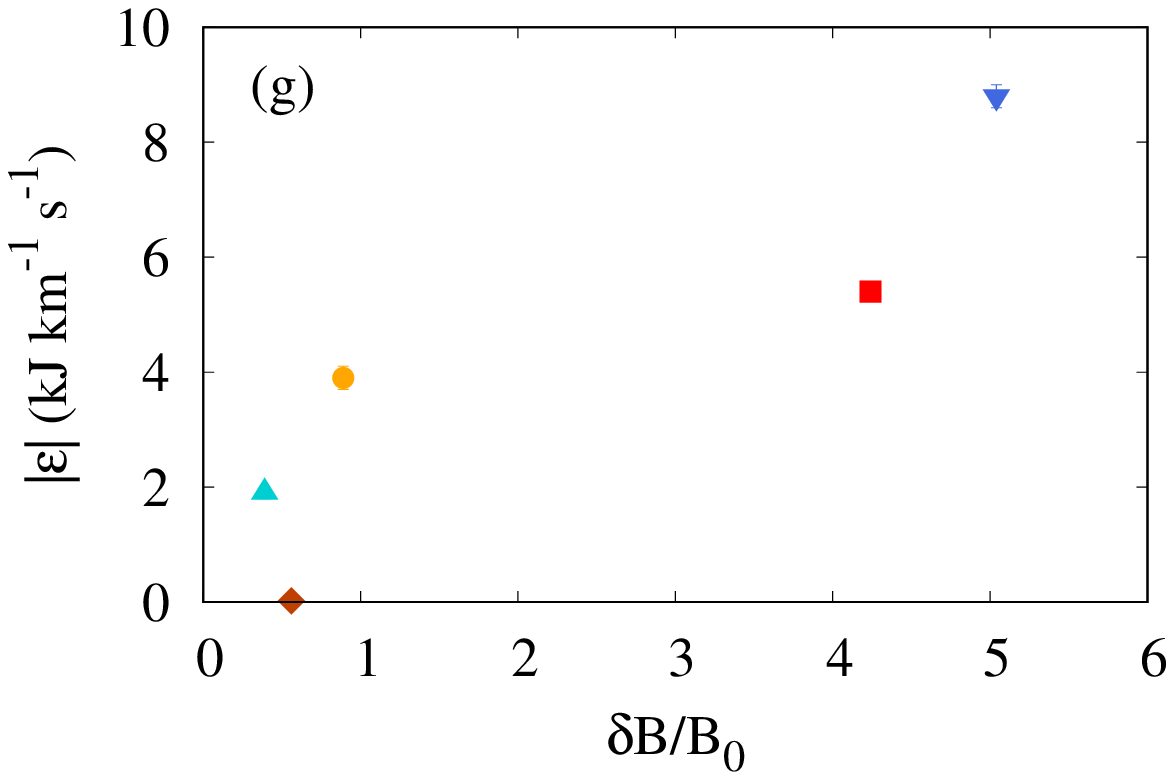}\includegraphics[width=0.33\textwidth]{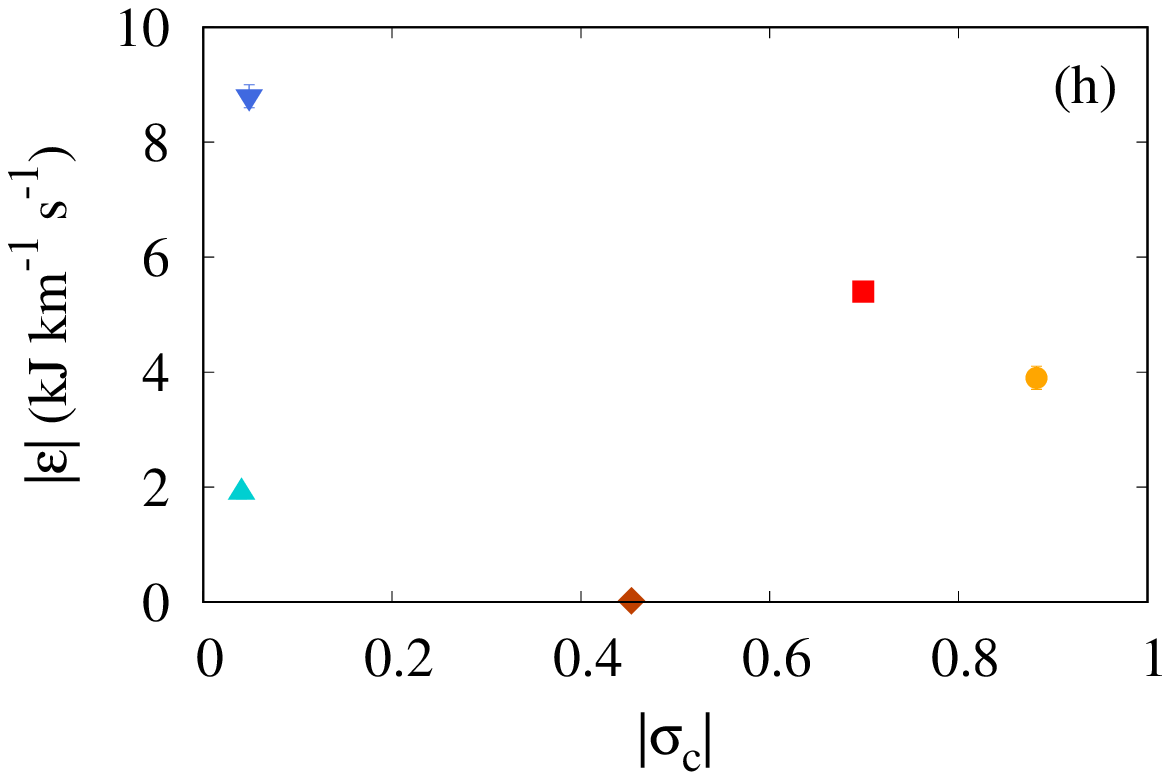}\includegraphics[width=0.33\textwidth]{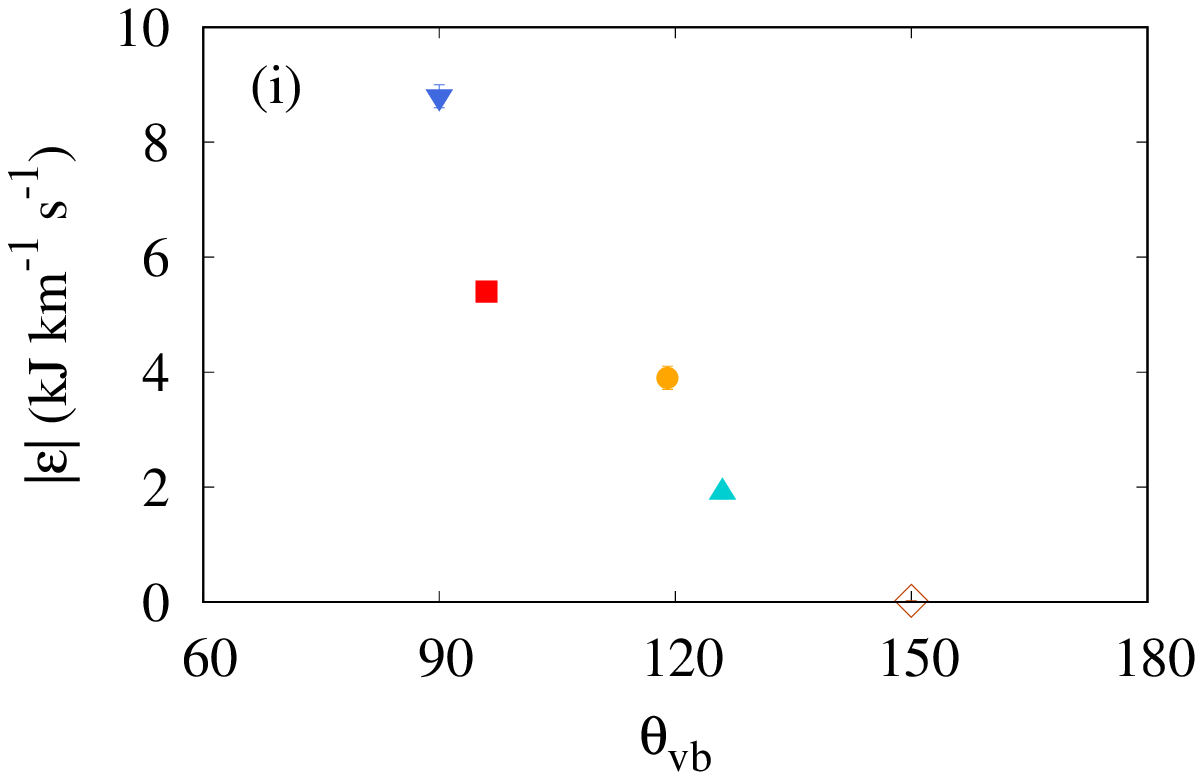}
    \includegraphics[width=0.33\textwidth]{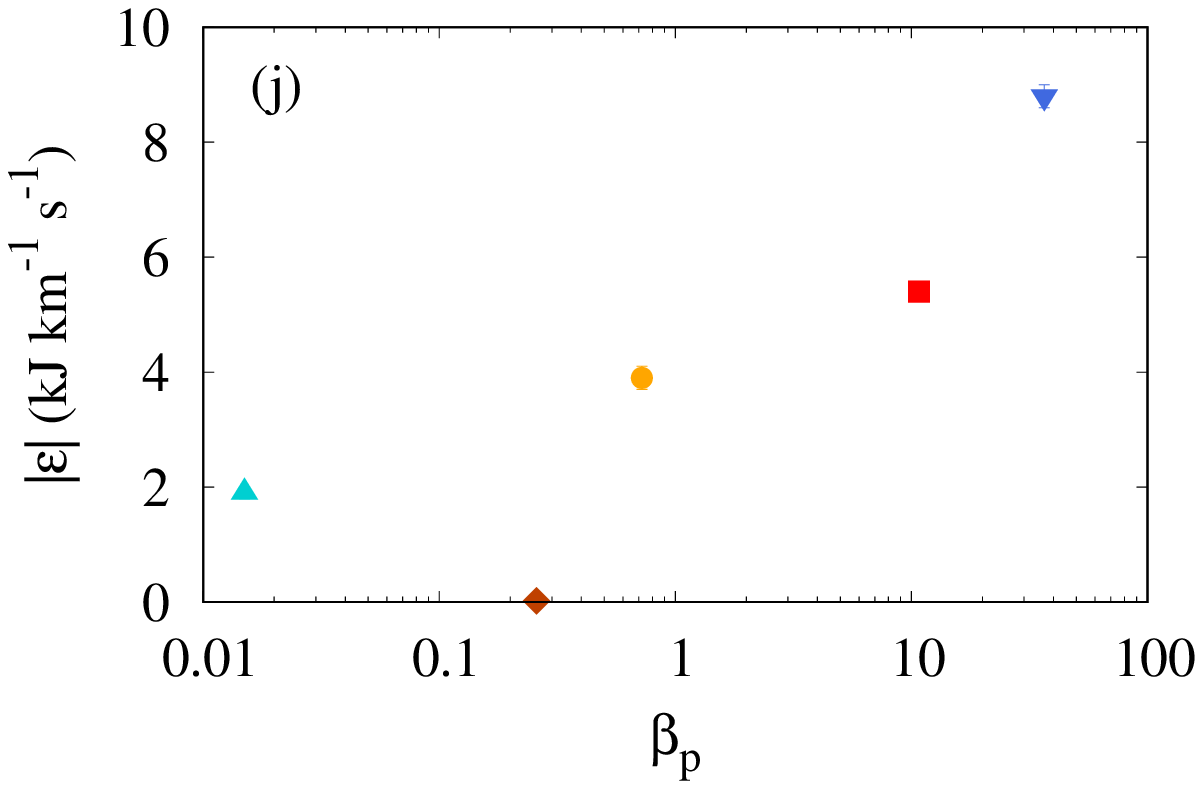}\includegraphics[width=0.33\textwidth]{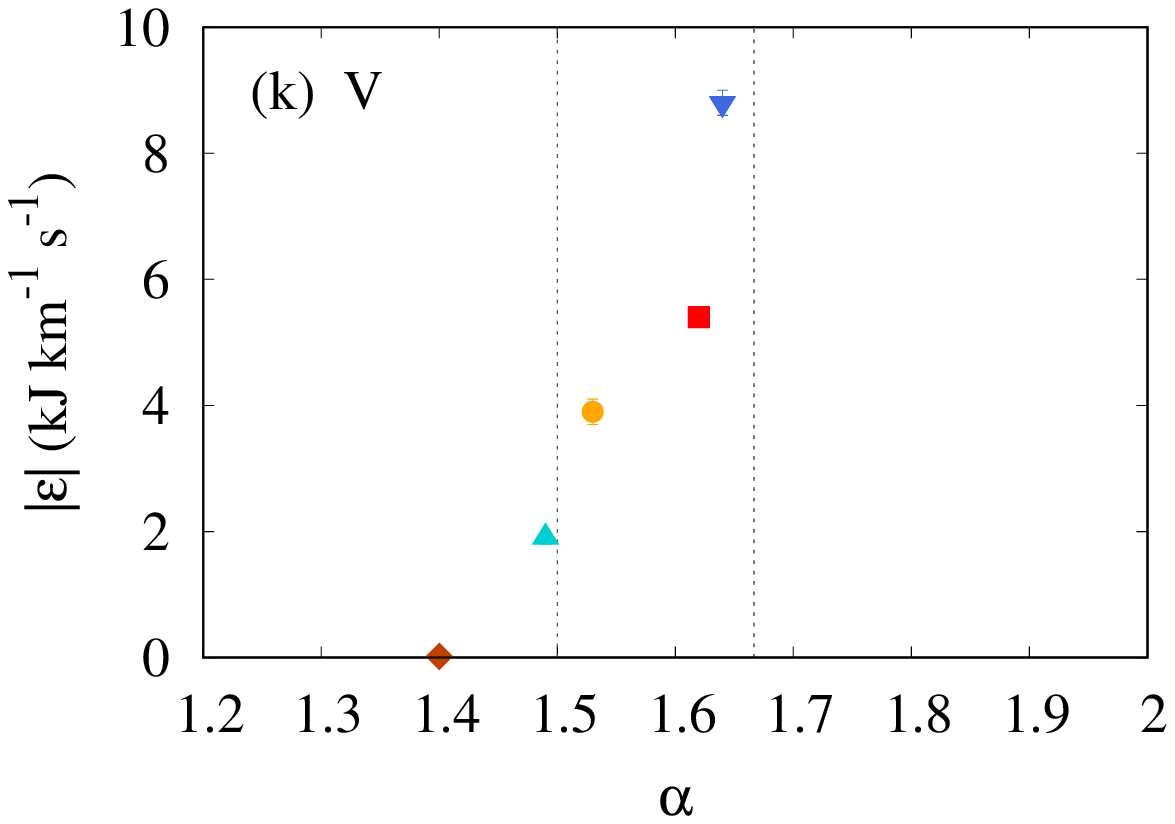}\includegraphics[width=0.33\textwidth]{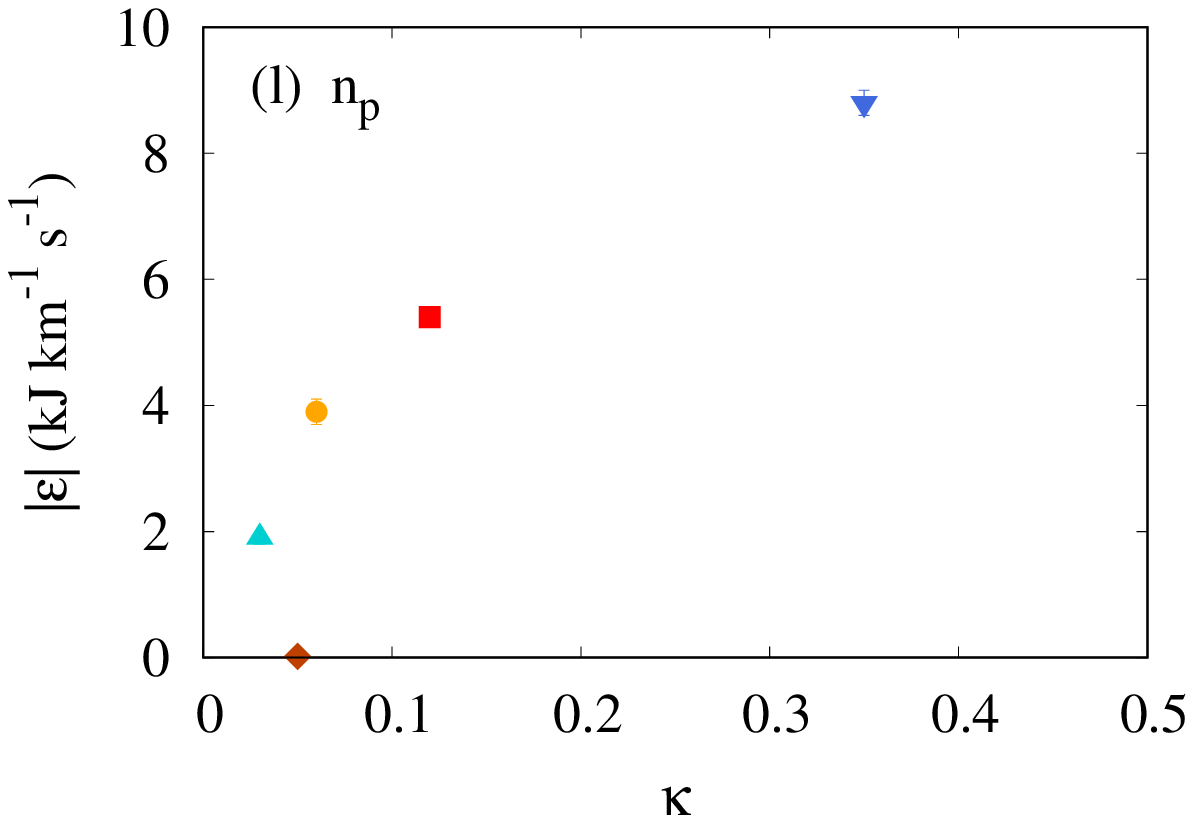}
    
    \caption{Panels $a$---$e$: third-order scaling law for the five samples. Linear fits are superposed in the relevant range, and the resulting energy transfer rate is given. Open symbols indicate positive $Y$ (and hence positive $\varepsilon$), full symbols indicate the opposite sign. 
    Panels $f$---$l$: absolute value of the energy transfer rate $|\varepsilon|$ versus various parameters: solar wind speed (panel $f$), level of magnetic fluctuations (panel $g$), normalized cross-helicity (panel $h$), angle between magnetic field and bulk flow (panel $i$), proton plasma $\beta$ (panel $j$), spectral index for velocity (panel $k$) and flatness exponent for proton density (panel $l$). All symbols are color-coded for different intervals, according to the labels in the bottom panel of Figure~\ref{fig:data}. In panel ($i$), for sample SW2 (open dark orange marker) the angle was transformed from 30$^\circ$ to 150$^\circ$ (assuming cylindrical symmetry around $B_0$) to improve visualization.}
    \label{fig:yaglom}
\end{figure}

For the evaluation of the PP law, we have estimated the time-dependent mixed third-order fluctuations of  Eq.~(\ref{Eq:PP}).
As evident from the top panel of Figure~\ref{fig:data}, in the regions under study the velocity is essentially radial. Therefore, the radial components of velocity and magnetic field can be used to compute the two-point longitudinal increments in the right-hand side of Eq.~(\ref{Eq:PP}). 
Some extreme values of the mixed third-order quantities were removed to achieve statistical convergence~\citep{Kiyani2006}. 
In the bottom panel of Figure~\ref{fig:data}, for each sample we show the Local Energy Transfer rate (LET), namely the unaveraged, timescale-normalized mixed third-order fluctuations $LET(\Delta t) = -3 [ \Delta v_L (|\Delta \mathbf{v}|^2+|\Delta \mathbf{b}|^2) - 2 \Delta b_L (\Delta \mathbf{v}\cdot\Delta \mathbf{b})]/(4 V_{sw} \Delta t)$, useful to determine the local contribution of the MHD field fluctuations to the global energy cascade~\citep{Sorriso-Valvo2018,Sorriso-Valvo2019}. 
The intermittent nature of the LET time series is evident, as well as the variety of fluctuation levels in the different regions. In particular, the trailing solar wind region SW-3 has extremely large fluctuations, while the SW-2 region has very weak LET. 
It is also clear that local contributions can be positive or negative, so that the actual global energy transfer results from the weak unbalance between the signed fluctuations.
The third-order moment, Eq.~(\ref{Eq:PP}), was then computed time-averaging over each interval. The resulting $Y(\Delta t)$ is plotted in panels $a$---$e$ of Figure~\ref{fig:yaglom} for the five intervals. 
Despite the relatively short samples and the inherently difficult observation of the third-order moments, the emergence of a linear scaling range is visible in most of the cases, with the sole exception of SW-2, where changes of sign in the inertial range break the PP law. 
This is the first observation of the PP law inside a CME, proving that the turbulence is fully developed even in the presence of the shock interaction~\citep{Kilpua2020}. 
The linear fit of the third-order moments as a function of the timescale $\Delta t$ provided the estimate of the mean energy transfer rate $\varepsilon$, whose absolute value is indicated in each panel of Figure~\ref{fig:yaglom}.
It is worth remarking that the sign of the third-order moment is, in principle, related to the direction of the energy flow in the cascade, so that positive (negative) $\varepsilon$ indicates energy flux towards small (large) scales. 
However, given the complex experimental conditions of solar wind turbulence, such correspondence is not always robust, and the actual meaningfulness of the sign is still an open research issue. 
Therefore, as become customary~\citep{Hadid2018}, in panels $f$---$l$ of Figure~\ref{fig:yaglom} we have used the magnitude $|\varepsilon|$ as an estimate of the amount of energy present in the turbulent cascade. For the sake of completeness, in panels $a$---$e$ the signed values of $\varepsilon$ were given, and we used open and full markers to indicate positive and negative values of $Y$, respectively. Considering the sign of the bulk velocity in the Taylor transformation as appearing in the $r.h.s.$ of Equation~\ref{Eq:PP}, these are associated with positive and negative energy transfer rate, respectively. With these caveats in mind, we note that the energy transfer rate sign was found positive in SW1, undefined in SW2, and negative in the remaining intervals.

According to the above analysis, the following features can be observed.
First of all, despite the fact that the largest LET fluctuations and wind speed are in the trailing plasma SW-3 (bottom panel of Figure~\ref{fig:data}), the highest energy transfer rate is found in the CME sheath. 
The enhancement of the energy transfer rate is likely due to local energy input from the shock interaction. 
Such energy is rapidly transported across scales by the nonlinear interactions, becoming incorporated in the turbulent cascade. 
The high energy transfer in the CME-SH region is also in agreement with the smaller cross-helicity. 
Indeed, the presence of unbalanced Alfv\'enic fluctuations reduces the efficiency of the cascade, as broadly observed in the solar wind~\citep{Smith2009}. 
Secondly, the high level of fluctuations observed in the SW-3 sample behind the CME, as compared with the preceding solar wind sample SW-1, does not correspond to a larger energy transfer rate. This may be due to the different solar wind conditions overall, and in particular to the larger cross-helicity in SW-3. 
The scatter plot in Fig.~\ref{fig:yaglom}$h$ reveals that larger energy transfer is measured for intervals with smaller cross-helicity, with the exceptions of the CME cloud, where the scaling is observed on smaller scales, and of the poorly developed SW2. 
Some ordering seems to be present with the solar wind speed (Figure~\ref{fig:yaglom}$f$), the faster wind having stronger energy transfer, again with the exception of the CME cloud. As it could be expected, a higher level of magnetic fluctuations $\delta B/B_0$ (where $B_0$ and $\delta B$ are the mean magnetic field and its {\it rms} value, both estimated averaging over each interval) is roughly associated with larger energy transfer rate, as shown in panel ($g$). Similarly, the proton plasma $\beta_p$ is also a good ordering parameter, with less magnetized plasma showing stronger energy cascade (panel $j$).
A better correlation is visible with the angle between the bulk flow and the mean magnetic field $\theta_{vb} = \cos^{-1}(\mathbf{v}\cdot\mathbf{B}/|\mathbf{v}||\mathbf{B}|)$, as shown in panel ($i$) (note that, for better visualization, for the SW2 interval the angle was transformed from 30$^\circ$ to 150$^\circ$, under the reasonable assumption of cylindrical symmetry of the turbulence around $B_0$). 
This suggests that intervals that are sampling the solar wind in the direction perpendicular to the mean field observe a stronger cascade, in agreement with the standard two-dimensional models of MHD turbulence~\citep{Narita2018}.
Finally, the spectral exponents of the velocity (panel $k$) and the flatness exponent of the density (panel $l$) both show positive correlation with $|\varepsilon|$, suggesting that faster solar wind and more compressive turbulence may both enhance the turbulent energy transfer. 
No clear correlation was observed for the other parameters and fields (not shown). 
%
%


\section{Conclusions} 
\label{sec:conclusions}

The properties of the turbulent cascade were studied inside a CME and in the surrounding solar wind using data measured by the Wind spacecraft on 12---20 July 2012. 
The analysis was performed using the scaling properties of the structure functions, which provided spectral and intermittency description. 
The fluctuation power and spectral index are enhanced in the CME sheath, in agreement with previous observations. 
The intermittency is comparable for all regions, with the remarkable exception of the density, which is intermittent only in the CME sheath where compressions are enhanced. 
The turbulent cascade was studied by means of the linear scaling of the mixed third-order fluctuation moments, or Politano-Pouquet law for MHD plasmas. 
Although the isotropic, incompressible version of the law was used, linear scaling was observed for the first time in the CME sheath region and in the CME magnetic cloud. 
The observation of the PP law provided a measure of the mean energy transfer rate associated with the turbulent cascade. This represents an approximate but important estimate of the turbulent energy available for activating and driving small-scale plasma processes. 
These may include various instabilities and wave-particle interaction mechanisms that eventually lead to energy dissipation, generation of waves, plasma heating and particle acceleration.
The observation provided evidence that the CME sheath is characterized by fully developed turbulence and enhanced energy transfer rate, which could be due to the energy injection at the shock interaction, to the strongly reduced unbalance of the Alfv\'enic fluctuations, and/or to the different sampling direction in the anisotropic solar wind turbulence.
In either cases, the interaction between shock and solar wind clearly plays a relevant role in enhancing the turbulent cascade and in increasing the associated energy transfer. 
This observation can be relevant for modeling the CME expansion and the role of the turbulent fluctuations in regulating the CME geoeffectiveness. 

\begin{acknowledgments}
LSV, EY and APD were funded by the Swedish Contingency Agency grant 2016-2102 and by SNSA grant 86/20. D.T. was partially supported by the Italian Space Agency (ASI) under contract 2018-30-HH.0.
\end{acknowledgments}

\end{document}